\documentclass[acmsmall]{acmart}

\usepackage[ruled,linesnumbered]{algorithm2e}

\SetCommentSty{mycommfont}

\newcommand{\myPar}[1]{\medskip\noindent\textbf{#1}.}
\newcommand{\myParNoSpace}[1]{\noindent\textbf{#1}.}

\usepackage{adjustbox}
\usepackage{amsmath}
\usepackage{listings}
\usepackage{tikz}
\usepackage{todonotes}
\usepackage{subcaption}
\usepackage{csquotes}

\usepackage[super]{nth}

\usepackage{pifont}
\newcommand{\cmark}{\ding{51}}
\newcommand{\xmark}{\ding{55}}

\usepackage{acronym}
\acrodef{API}{Application Programming Interface}
\acrodef{APR}{Automatic Program Repair}
\acrodef{AUG}{API Usage Graph}

\usepackage{xspace}
\newcommand{\tool}{ChaRLI\xspace}

\newcommand{\ovgu}{ Otto-von-Guericke University Magdeburg}
\newcommand{\rub}{ Ruhr-University Bochum}

\definecolor{ckeywordcolor}{RGB}{127,0,85}
\definecolor{cstringcolor}{RGB}{42,0,255}
\definecolor{ccommentcolor}{RGB}{63,127,95}
\definecolor{addcolor}{RGB}{34,136,51}
\definecolor{delcolor}{RGB}{238,102,119}

\newcommand*\circled[1]{\tikz[baseline=(char.base)]{
		\node[shape=circle,draw,inner sep=0.75pt] (char) {\textbf{#1}};}}

\graphicspath{{./}}

\begin{document}

\title{Automated Change Rule Inference for Distance-Based API Misuse Detection}

\settopmatter{authorsperrow=4}

\author{Sebastian Nielebock}
\email{sebastian.nielebock@ovgu.de}
\orcid{0000-0002-0147-3526}
\affiliation{%
	\ovgu
    \country{Germany} 
}

\author{Paul Blockhaus}
\email{paul.blockhaus@ovgu.de}
\orcid{0000-0001-6910-9475}
\affiliation{%
	\ovgu
    \country{Germany} 
}

\author{Jacob Krüger}
\email{jacob.krueger@rub.de}
\orcid{0000-0002-0283-248X}
\affiliation{%
	\rub
    \country{Germany} 
}

\author{Frank Ortmeier}
\email{frank.ortmeier@ovgu.de}
\orcid{0000-0001-6186-4142}
\affiliation{%
	\ovgu
	\country{Germany} 
}

\renewcommand{\shortauthors}{Nielebock et al.}

\begin{abstract}
  Developers build on Application Programming Interfaces (APIs) to reuse existing functionalities of code libraries.
Despite the benefits of reusing established libraries (e.g., time savings, high quality), developers may diverge from the API's intended usage; potentially causing bugs or, more specifically, \emph{API misuses}.
Recent research focuses on developing techniques to automatically detect API misuses, but many suffer from a high false-positive rate. 
In this article, we improve on this situation by proposing \tool{} (Change RuLe Inference), a technique for automatically inferring \emph{change rules} from developers' fixes of API misuses based on API Usage Graphs (AUGs). 
By subsequently applying graph-distance algorithms, we use change rules to discriminate API misuses from correct usages. 
This allows developers to reuse others' fixes of an API misuse at other code locations in the same or another project.
We evaluated the ability of change rules to detect API misuses based on three datasets and found that the best mean relative precision (i.e., for testable usages) ranges from 77.1\,\% to 96.1\,\% while the mean recall ranges from 0.007\,\% to 17.7\,\% for individual change rules. 
These results underpin that \tool{} and our misuse detection are helpful complements to existing API misuse detectors.
\end{abstract}

\ccsdesc[500]{Software and its engineering~Error handling and recovery}
\ccsdesc[500]{Software and its engineering~Automatic programming}
\ccsdesc[300]{Software and its engineering~Software version control}

\keywords{API Misuse, Misuse Detection, Change Rules, Cooperation}

\maketitle

\section{Introduction}

An Application Programming Interface (API) allows client developers to call the functionalities of another programming library within their own application, enabling software reuse of established code.
However, developers may not be fully aware of the correct usage of an API, for instance, which mandatory function calls are required in what order to use an API interface correctly.
As a consequence, developers may misuse the API, potentially causing bugs, faulty behavior, or unexpected outcomes in their own application.
We refer to such cases as \emph{API misuses}.
Unfortunately, API misuses are prevalent in software development.
For instance, \citet{Zhong2015} have shown that half of the bug fixes in five open-source Apache projects required at least one API-specific change.
Moreover, API misuses may cause severe bugs, for instance, security issues caused by falsely applied cryptography APIs~\cite{Nadi2016,Oliveira2018}.

To mitigate such problems, a large research community is working on techniques for detecting API misuses~\cite{Amann2018a,Amann2019,Kang2021,Nielebock2021a}.
The common idea is to infer specifications that describe the correct usage of an API from source code or other documents (e.g., API documentation) and to subsequently detect violations of these specifications.
Specifications are represented as state automata~\cite{Ammons2002}, dynamic invariants~\cite{Ernst2001}, temporal specifications~\cite{Wasylkowski2007}, API usage patterns based on frequent pattern mining~\cite{Weimer2005, Zhong2009}, or machine-learned probability distributions~\cite{Allamanis2014, Murali2017}.
However, a major issue of existing specification-based misuse detection techniques is the large number of irrelevant or alternative specifications causing false alarms (i.e., false positives)~\cite{Goues2012, Amann2019a}.
Such false alarms hamper the practical adoption of API misuse detectors---similarly to static code analyzers~\cite{Imtiaz2019StaticAnalysisActing,Imtiaz2019ChallengesStaticAnalysisPractice}.

To tackle the problem of reporting many false alarms, we have introduced the idea of reusing existing knowledge from fixed API misuses to detect similar misuses in other projects with a lower false-positive rate~\cite{Nielebock2020}.
Essentially, we build on the idea that a fixed misuse comprises detailed information about the misuse, its fix, and the corrective changes---such as the context of the API misuse and how a real developer fixed that API misuse in that specific context.
We have developed a technique that uses the change information of commits to semi-automatically infer \emph{correction rules} for misuses, which can ideally be reused to detect and eventually fix similar misuses at other code locations.
Subsequently, we empirically analyzed to what extent we can employ distance-based metrics to compare correction rules to API misuses and correct API usages~\cite{Nielebock2021b}.
Unfortunately, we identified two major challenges hampering our idea of detecting API misuses.
First, generating reliable correction rules required extensive manual effort, for instance, to identify and denote the misused API.
Second, the distance-based comparisons achieved low precision values compared to advanced specification-based techniques, such as MUDetect~\cite{Amann2019} or ALP~\cite{Kang2021}.

In this article, we report on our advancements with which we tackled these two challenges.
To this end, we propose \tool{} (Change RuLe Inference), a novel technique for inferring \emph{change rules}---a generalization of the correction rules we used before.
Additionally, we evaluated to what extent the inferred change rules can help to reliably discriminate API misuses from correct API usages.
More detailed, we contribute the following in this article:
\begin{itemize}
	\item We propose \tool{}, a technique that, except for the manually provided commit hash and method declaration containing the fixed API misuse, can automatically infer change rules.
	
	\item We improve distance-based metrics to detect API misuses based on inferred change rules.
	
	\item We report an extensive evaluation of \tool{} and our distance based misuse detection on three different datasets: MUBench~\cite{Amann2016}, AU500~\cite{Kang2021}, and AndroidCompass~\cite{Nielebock2021}.

	\item We publish all artifacts related to this article in an open-access repository.\footnote{\url{https://doi.org/10.5281/zenodo.6598541}\label{fn:replication}}
\end{itemize}
Our contributions can help practitioners to detect API misuses cooperatively by one developer proposing a change rule that others can apply for the same API in their projects (e.g., after a breaking change in an API).
Moreover, \tool{} provides a foundation for developing and integrating tools to facilitate the inference of change rules.
For researchers, our contributions define indicators on how to improve misuse detection, providing a reusable and extensible foundation for this purpose.
\section{Background}\label{sec:background}

In this section, we briefly introduce the basic concepts of AUGs, correction rules, and distance-based API misuse detection.

\subsection{API Usage Graphs}

An AUG is a directed labeled multi-graph describing the data and control flow of an intra-procedural API usage.
The concepts of AUGs have been developed by \citet{Amann2019a} to represent API specifications for more effective API-misuse detection.
In our previous works, we used AUGs to mine specifications~\cite{Nielebock2021a} and to produce correction rules~\cite{Nielebock2020, Nielebock2021b}.
Note that AUGs target only the Java programming language.

\begin{figure}
	\begin{subfigure}[b]{0.495\linewidth}
		\centering
\begin{lstlisting}
package pkg.at.some.loc;

import from.another.place.Foo;
import from.another.place.Bar;
import from.another.place.Baz;

public class AUGSample {
  Foo fooObj = new Foo(42,"text");

  public Integer computeSomething(Bar barObj) {
    Baz bazObj = new Baz(this.fooObj);
-   bazObj.doSomething(barObj);
+   if(bazObj.hasCharacteristic()){
+       bazObj.doSomething(barObj);
+    }
    return bazObj.getResult();
  }
}
\end{lstlisting}
        \caption{Source code changes.\label{lst:code}}
	\end{subfigure}
	\begin{subfigure}[b]{0.495\linewidth}
		\centering\includegraphics[width=0.9\textwidth]{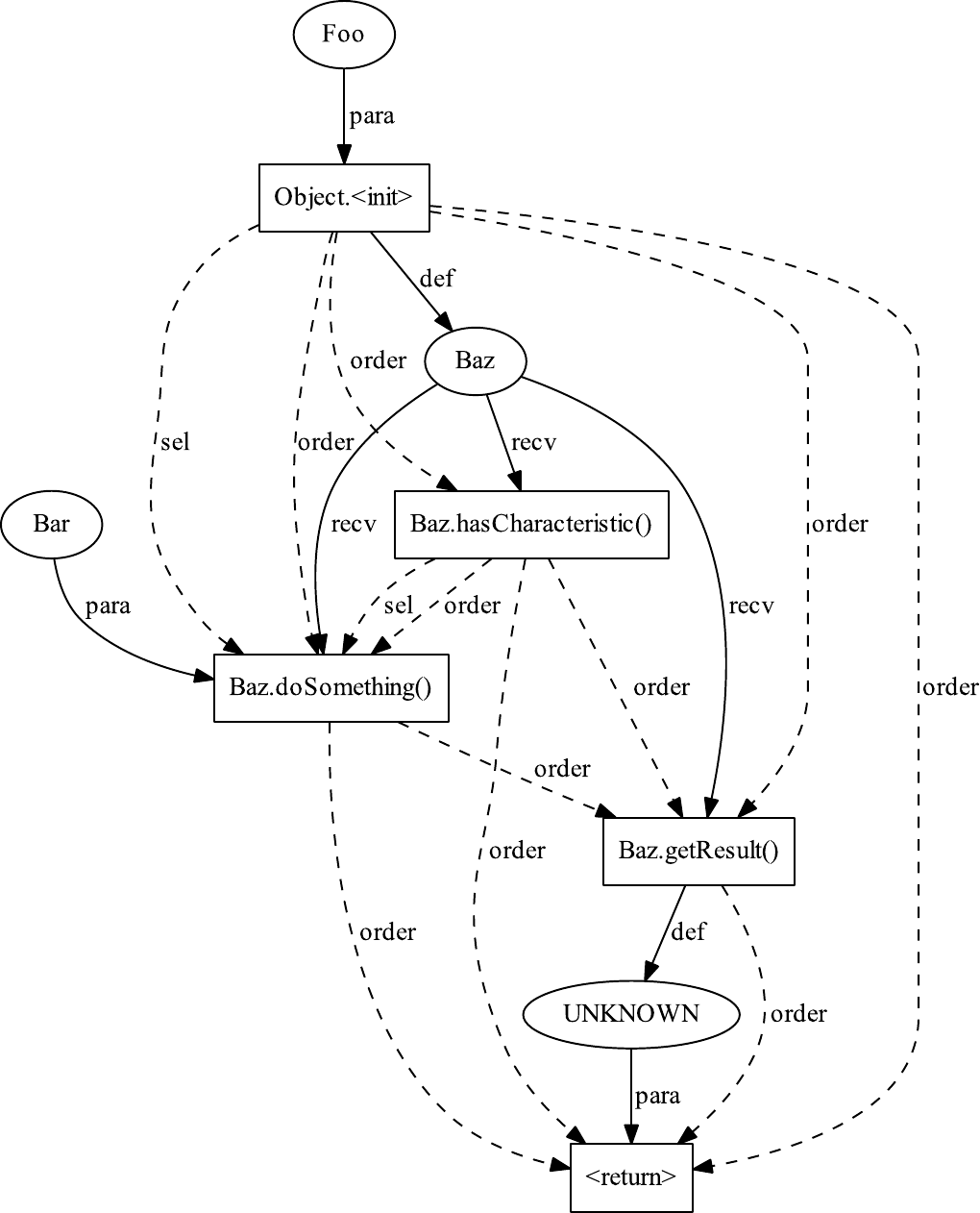}
        \caption{AUG of the changed code.\label{fig:aug}}

	\end{subfigure}
	\caption{Fix of a fictive API misuse in an application by adding an additional check to \lstinline[basicstyle=\ttfamily]|bazObj| (\autoref{lst:code}, Lines 12--15) and the AUG of the fixed code version (\autoref{fig:aug}).}
	\label{fig:example}
\end{figure}


For simplicity, we introduce AUGs based on the example we used in our previous work~\cite{Nielebock2021b}, and which we display in \autoref{fig:example}.
In \autoref{lst:code}, we show a code excerpt that a developer has changed (i.e., by adding the \lstinline[basicstyle=\ttfamily]|hasCharacteristic()|-check from Lines 12--15). 
The AUG of the changed code in \autoref{fig:aug} builds on the code's Abstract Syntax Tree (AST) enhanced by heuristically obtained control and data flow edges. 
Moreover, type resolution is achieved by attaching additional source code (e.g., by providing the \texttt{jar} of a certain library) to the construction process.

An AUG consists of two main types of nodes: data nodes (displayed as ellipses in \autoref{fig:aug}) and action nodes (displayed as rectangles in \autoref{fig:aug}). 
Data nodes represent object instances, which are labeled either with their respective data types or, in the case of constants, their respective value (e.g., the text in case of a \lstinline[basicstyle=\ttfamily]|String| constant). 
If a datatype cannot be resolved, the data node is labeled with \lstinline[basicstyle=\ttfamily]|UNKNOWN| (cf. \autoref{fig:aug}). 
Action nodes represent method calls (e.g., \lstinline[basicstyle=\ttfamily]|Baz.doSomething()|) or special control-flow statements (e.g., \lstinline[basicstyle=\ttfamily]|<return>|). 
These generic node types can be refined in the actual implementation of an AUG.

An AUG involves two primary types of edges: control-flow and data-flow edges. 
Control-flow edges are depicted as dashed arrows and labeled with their respective sub-type. For instance, \lstinline[basicstyle=\ttfamily]|sel| denotes a selective edge representing the changed control flow caused by \lstinline[basicstyle=\ttfamily]|if|-statements. 
Another example are \lstinline[basicstyle=\ttfamily]|order|-edges, which denote the heuristically (i.e., transitive closure) obtained order of statements. 
Data-flow edges describe the data flow between nodes and are represented as solid arrows. 
Examples of such edges are \lstinline[basicstyle=\ttfamily]|recv|-edges, which represent that a certain action (i.e., target node) is called from the incoming data node, or \lstinline[basicstyle=\ttfamily]|para|-edges, which represent that the incoming data node is used as a parameter. 

\begin{figure}
	\centering\includegraphics[width=0.75\textwidth]{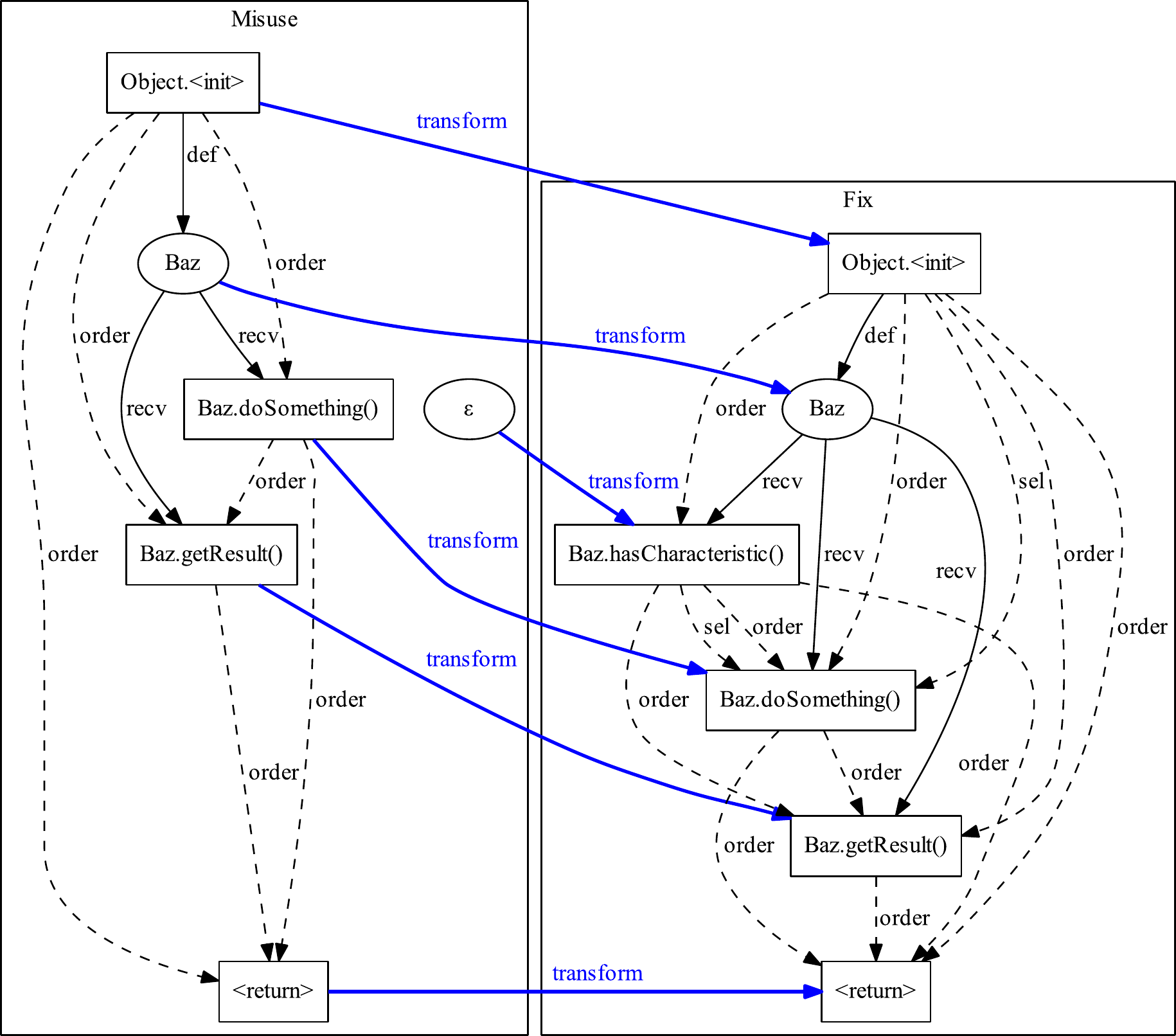}
	\caption{Correction rule for the fix of the API misuse in \autoref{lst:code}, adapted from our previous work~\cite{Nielebock2021b}.}
	\label{fig:corr_rule}
\end{figure}

\subsection{Correction Rules\label{ssec:background_correctionrule}}

In our previous work~\cite{Nielebock2020,Nielebock2021a}, we proposed so-called \emph{correction rules} to detect API misuses. 
A correction rule represents the changes between two subsequent versions of an AUG that are intended to correct an API misuse. 
Particularly, each rule should represent the minimal changes between the nodes of the misuse AUG and the corrected AUG, depicted as \lstinline[basicstyle=\ttfamily]|transform|-edges. 
In this context, added or deleted nodes are represented as special $\epsilon$-nodes in the misuse or fix part of a correction rule (i.e., representing \enquote{holes} in the respective other AUG). 
We show the correction rule for the changes in \autoref{lst:code} in \autoref{fig:corr_rule}.
Note that this rule simplifies the respective misuse and fix AUGs by removing nodes that do not change, meaning nodes for which neither the label nor the in- or outgoing edges change. 
Therefore, the correction rule depicted in \autoref{fig:corr_rule} contains neither the data nodes of the objects \lstinline[basicstyle=\ttfamily]|Foo| and \lstinline[basicstyle=\ttfamily]|Bar| nor their respective edges, which are present in the original AUG (cf. \autoref{fig:aug}).

The minimal number of changes between two AUGs can be computed using the minimal Graph Edit Distance (GED), which identifies the minimal edit costs in terms of change operations to transform one graph into another (i.e., by adding, deleting, or relabeling nodes and edges). 
However, GED computation is known to be NP-hard~\cite{Blumenthal2018}. 
As a result, we~\cite{Nielebock2020} previously applied an approximation technique to obtain an almost minimal GED. Particularly, we transform the two AUGs into a bipartite mapping of the nodes and apply the Kuhn-Munkres algorithm~\cite{Munkres1957} to obtain a sufficiently minimal mapping. 
This reduces the problem's complexity. 
Since we do not change this part of our algorithm, we describe its details in~\autoref{ssec:charli}.
Moreover, to reduce the number of nodes involved in a correction rule, we filter those nodes that refer to a certain type (i.e., class). 
Particularly, while our original idea focused on a manual selection of relevant classes for an API misuse, we generalized this in our subsequent work~\cite{Nielebock2021b}. We automatically removed nodes not related to any imported class or a certain subset of the imported classes (e.g., \lstinline[basicstyle=\ttfamily]|android.*|) of the analyzed source file.

\subsection{Distance-Based API Misuse Detection}\label{ssec:background_dist}

Using correction rules, we proposed to detect API misuses based on graph-distance algorithms~\cite{Nielebock2021b}. 
Particularly, our idea is to compute distance values between an API usage (by inferring its AUG) and i) the respective misuse part as well as ii) the fixed part of the correction rule. 
In case the usage is more similar (i.e., has a smaller distance) to the misuse part than to the fixed part, we determine the API usage as an API misuse.

More formally, we describe a correction rule as $aug_{rm} \rightarrow aug_{rc}$ where $aug_{rm}$ (\textbf{r}ule \textbf{m}isuse) and $aug_{rc}$ (\textbf{r}ule \textbf{c}orrection) are the respective misuse and fix AUGs. Moreover, we define $dist$ to be a relative distance function $dist: AUG \times AUG \rightarrow [0,1]$, where $0$ denotes equality and $1$ maximum inequality between two AUGs. Having an API usage represented as AUG $aug_{x}$, we say $aug_{x}$ is a misuse according to the above mentioned correction rule and distance function if
\begin{equation}\label{eq:misuse-detect}
	dist(aug_{rm},aug_{x})<dist(aug_{rc},aug_{x})
\end{equation}
Additionally, because some rules may only match this condition by chance, we checked whether correction rules are \emph{applicable} for misuse detection. 
This means that, according to a set of known correct API usages, the correct part of the rule has a smaller mean distance to such API usages than to a set of known API misuses. 
Similarly, the misuse part of a correction rule should have a smaller distance to the known API misuses than to the known correct API usages. 
The detailed computation of this applicability check can be found in our previous work~\cite{Nielebock2021b}. 
We found that this check together with different graph-distance algorithms (e.g., L1-norm and cosine similarity based on the Exas vectors proposed by \citet{Nguyen2009}) are insufficient for detecting API misuses.
Reflecting on these outcomes, we hypothesized how to improve the distance function and enable API misuse detection via AUGs, which we discuss in \autoref{sec:charli_and_dist}.
\section{API Misuse Detection Using Change Rules}\label{sec:charli_and_dist}

\begin{figure}
	\includegraphics*[width=0.7\textwidth]{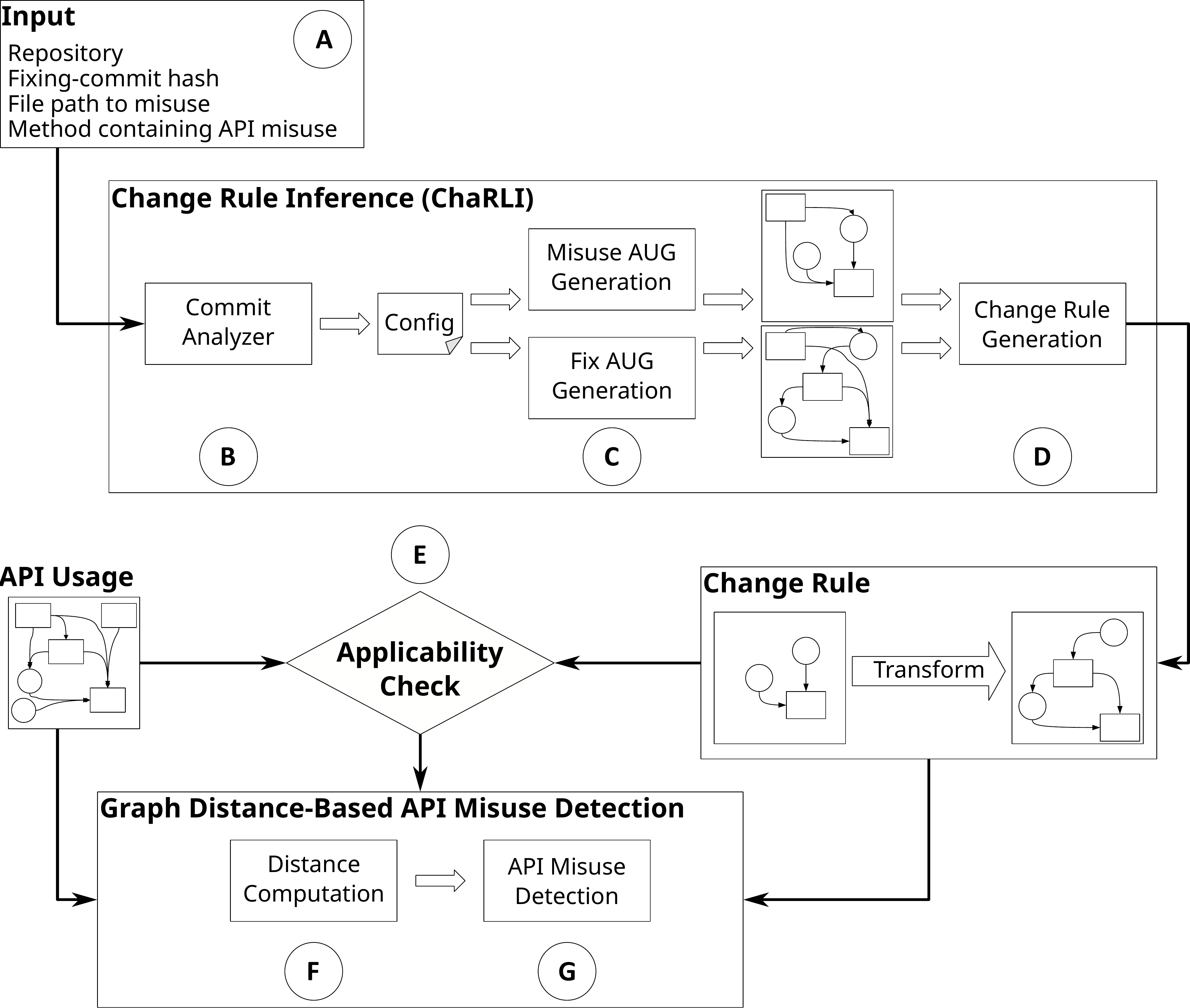}
	\caption{Overview of our change-rule inference and graph-distance-based API misuse detection.}\label{fig:charli_detect_process}
\end{figure}

In this section, we start with an overview of our concept for detecting API misuses via change rules (cf. \autoref{ssec:process}).
Then, we introduce the individual parts of our concept. 
First, we present \tool{} by describing a validation of the issues we identified for our correction rules and describing our consequent improvements (cf. \autoref{ssec:charli}). 
Second, we present how we check the applicability of a change rule to a potential API misuse (cf. \autoref{ssec:applicability}).
Finally, we present the graph-distance-based API misuse detection---especially our improvements of Exas vectors (cf. \autoref{ssec:dist}). 

\subsection{Conceptual Overview}\label{ssec:process}

We depict the conceptual overview of our techniques in \autoref{fig:charli_detect_process}, involving \tool as well as the graph-distance-based API misuse detection. 
This process starts within the project in which one developer identified and fixed an API misuse. 
As input for \tool, we need the repository URI of the project, the commit hash of the fixing change, and the path to the source file as well as to the method declaration of the fixed method ({\footnotesize\circled{A}}). 
Then, \tool analyzes the commit ({\footnotesize\circled{B}}) to produce a configuration for the AUG generation. 
Using this configuration, \tool produces two AUGs, one representing the misused and the other representing the fixed code version ({\footnotesize\circled{C}}). 
Lastly, \tool automatically detects the differences between the two AUGs to generate the change rule ({\footnotesize\circled{D}}).

Afterward, we use the change rule to detect the API misuse represented by this rule in other API usages, which may reside in the same or another project. 
For the misuse detection, we first perform an applicability check of the rule against the API usage ({\footnotesize\circled{E}}). 
If the rule is applicable to the API usage, we use the change rule to perform the graph distance-based API misuse detection. 
This final part of our process involves computing the distance between the API usage and the misuse as well as the fixed part of the change rule ({\footnotesize\circled{F}}); and the subsequent misuse detection ({\footnotesize\circled{G}}). 
Next, we describe the individual steps of this process in more detail.

\subsection{Change Rule Inference}\label{ssec:charli}

We begin this section with a validation of our previous correction rules (cf. Section~\ref{ssec:background_correctionrule}). This way, we identified improvements for generating our novel change rules, for which we describe the generation steps afterward.

\looseness=-1
\myPar{Validation of Correction Rules}
The goal of our manual validation of correction rules~\cite{Nielebock2021b} is to identify hypotheses to improve the precision for API misuses detection. 
For this purpose, the first three authors independently analyzed 50 randomly selected API misuses from the MUBench dataset~\cite{Amann2016} and the respective correction rules.
They compared each rule to the changes made in the fix commit to check whether the rule correctly represents the steps needed to perform the fix. 
Particularly, we assessed the rules based on the following three criteria:
\begin{enumerate}
	\item \emph{Completeness}: All necessary API methods (i.e., AUG action nodes) of the API misuse fix are present.
	\item \emph{Arrangement}: All necessary API methods (i.e., AUG action nodes) are correctly arranged concerning the control flow.
	\item \emph{Data}: All required parameters and return values (i.e., AUG data nodes) of the necessary API methods are part of the change rule.
\end{enumerate}
We agreed on these minimal criteria to denote a rule as valid and assessed each criterion individually as \emph{yes}, \emph{no}, or \emph{do not know}. 
Moreover, we justified each decision and codified additional flaws we noticed. 
So, we could also identify issues in the rules going beyond those three criteria.
In the end, we discussed the individual assessments to derive agreement on each of the fifty correction rules among the three authors.

After this validation, we summarized the results by denoting a criterion of a rule as \emph{yes} if all three assessors voted \emph{yes}. 
We denoted all other cases as \emph{no}. 
Overall, Cohens $\kappa$~\cite{Cohen1960} indicates an agreement for completeness of $\kappa\approx0.64$, for arrangement of $\kappa\approx0.57$, and for data of $\kappa\approx0.26$. 
According to the criteria defined by \citet{Landis1977}, these scores refer to a substantial (for completeness), a moderate (for arrangement), and a fair agreement (for data). 
Please note that these agreement strengths are debatable even though they are commonly used to measure agreement, particularly because \citeauthor{Landis1977} denote their criteria as \enquote{arbitrary.} 

\begin{table}
	\caption{Summary of our manual validation of 50 randomly produced correction rules from MUBench\label{tab:rule_validation}.}
	\begin{tabular}{lrrr}
		\hline
		\textbf{Decision} & \textbf{\#Completeness} & \textbf{\#Arrangement} & \textbf{\#Data}\\
		\hline
		\textbf{yes} & 18 & 14 & 0\\		
		\textbf{no} & 32 & 36 & 50\\
		\hline		
	\end{tabular}
\end{table}

In \autoref{tab:rule_validation}, we depict the results of our manual validation. 
We can see that we denoted only a minority of 18 rules as complete and only 14 as correctly arranged. 
None of the rules have been assessed as correctly containing parameters or return values by all assessors. 
Through our codification, we denoted potential reasons for these results from which we derived the following three challenges:
\begin{description}
	\item[Challenge 1:] Missing data nodes were mainly a consequence of how we denoted changed nodes. 
	A node changes if it is relabeled or if its incoming and outgoing edges change. 
	However, this usually does not impact parameters and result object nodes. 
	For instance, for the AUG in \autoref{fig:aug}, the data nodes \lstinline[basicstyle=\ttfamily]|Foo|, \lstinline[basicstyle=\ttfamily]|Bar|, and \lstinline[basicstyle=\ttfamily]|UNKNOWN| are only connected to their neighboring action nodes via data flow edges (i.e., \lstinline[basicstyle=\ttfamily]|def| and \lstinline[basicstyle=\ttfamily]|para|), while the action nodes have some additional control-flow edges, such as \lstinline[basicstyle=\ttfamily]|order|-edges. 
	Since adding a method call only impacts the control-flow edges, the mentioned data nodes do not change according to our criteria, and thus do not occur in the correction rule (cf. \autoref{fig:corr_rule}). 
	
	\item[Challenge 2:] We found a similar issue for some missing action nodes connected via a \lstinline[basicstyle=\ttfamily]|finally| control-flow edges. 
	Such an edge describes the control flow to the \lstinline[basicstyle=\ttfamily]|finally|-block in Java, which executes statements (e.g., tidy up statements) even in the case of an exception. 
	
	\looseness=-1
	\item[Challenge 3:] During our manual inspection, we noticed that the conservative addition of \lstinline[basicstyle=\ttfamily]|order|-edges as a transitive closure increased the number of such edges drastically (e.g., in an extreme case a correction rule contained roughly 3,000 order edges)---making it practically impossible to manually inspect that rule. 
	Moreover, this huge amount of edges may have negative consequences on the distance computation. 
	For instance, when applying Exas vectors (cf. \autoref{ssec:background_dist} and \autoref{ssec:dist}), the amount of edges also increases the number of features of the respective vector. 
	We found many \lstinline[basicstyle=\ttfamily]|order|-edges in our examples to be irrelevant because there exist other connections between the same nodes (e.g., via data-flow edges). 
\end{description}
Next, we detail the individual steps of \tool{} and how we tackled these challenges.

\myPar{Commit Analyzer ({\footnotesize\circled{B}})} 
The commit analyzer automatically creates a configuration file based on which the subsequent AUG generation produces the respective misuse and fix AUG. 
For this purpose, the analyzer requires the location of a misuse (i.e., the source file and the method declaration containing the misuse) as well as the commit and the URI of the repository ({\footnotesize\circled{A}}). 
Currently, we only support git repositories, because they are prevalent among open-source projects (e.g., on GitHub or BitBucket). 
The analyzer additionally identifies the source-code root path of the project to improve type inference. 
This is done via the respective \lstinline[basicstyle=\ttfamily]|package| declaration of the source file containing the misuse as well as its file path (e.g., if the source file path is \lstinline[basicstyle=\ttfamily]|/home/foo/de/bar/project/pkg_a/subpkg_d/File.java| and the \lstinline[basicstyle=\ttfamily]|package| declaration would be \lstinline[basicstyle=\ttfamily]|de.bar.project.pkg_a.subpkg_d| then the source-root path is \lstinline[basicstyle=\ttfamily]|/home/foo/|). 
Additionally, the developer may add paths to libraries as \lstinline[basicstyle=\ttfamily]|*.jar|-files.

\myPar{AUG Generation ({\footnotesize\circled{C}})}
Using the configuration file, \tool{} checks out the version after the fixing commit (i.e., \lstinline[basicstyle=\ttfamily]|fixingCommitHash|) as fix and the version before the fixing commit (i.e., \lstinline[basicstyle=\ttfamily]|fixingCommitHash~1|) as misuse.  
For each commit version, \tool{} generates for the specified method declaration in the given source file the respective AUG. 
Similar to our previous work~\cite{Nielebock2021b}, we also store for each node the respective API type, if available. 
For instance, node \lstinline[basicstyle=\ttfamily]|Baz| in \autoref{fig:aug} internally stores the corresponding type \lstinline[basicstyle=\ttfamily]|java.lang.Object|. 
In case the automated type inference could not determine the type, it is set to an empty \lstinline[basicstyle=\ttfamily]|String|.

\myPar{Change Rule Generation ({\footnotesize\circled{D}})}
We depict our algorithm for generating change rules in \autoref{alg:change_rule}. 
For clarity, we omit the straightforward computations and simplified implementation details. Please refer to our replication package\textsuperscript{\ref{fn:replication}} for the detailed implementation.

\begin{algorithm}
	\DontPrintSemicolon
	\caption{Change Rule Generation\label{alg:change_rule}}
	\KwIn{$aug_m, aug_c$}
	\KwResult{Change Rule: $(aug'_m \rightarrow aug'_c)$}
	\While{$\lvert aug_m.nodes\rvert \neq \lvert aug_c.nodes\rvert$}{add $\epsilon$-nodes to the AUG with fewer nodes}
	$bipartite.nodes \gets aug_m.nodes \cup aug_c.nodes$\;
	$bipartite.edges \gets \{(v_m, v_c, label: cost(v_m, v_c))$ with $v_m \in aug_m.nodes \land v_c \in aug_c.nodes\}$\;
	$ minMap \gets kuhn\_munkres(bipartite)$\;
	$aug'_m, aug'_c \gets $ create AUGs with nodes from $minMap$ with $costs>0$ as well as their corresponding edges from $aug_m$ and $aug_c$\;
	$single\_hop\_node\_pairs \gets $ all action-node pairs from $minMap$ that are connected with an incoming data-flow or \lstinline[basicstyle=\ttfamily]|finally| edge (in $aug_m$ and $aug_c$, respectively) to one of the nodes in $aug'_m$ or $aug'_c$\;
	\ForEach{$(sh\_node_m,sh\_node_c)\in single\_hop\_node\_pairs$}{
	Add $sh\_node_m$ with its corresponding edges (from $aug_m$) to $aug'_m$\;
	Add $sh\_node_c$ with its corresponding edges (from $aug_c$) to $aug'_c$\;
	}	
	$post\_single\_hop\_node\_pairs \gets \emptyset$\;
	\ForEach{$(sh\_node_m,sh\_node_c)\in single\_hop\_node\_pairs$}{
	$post\_single\_hop\_node\_pairs \gets post\_single\_hop\_node\_pairs\ \cup$ all data-node pairs from $minMap$ that are connected with an outgoing data-flow edge (in $aug_m$ and $aug_c$, respectively) to one of the nodes $sh\_node_m$ or $sh\_node_c$\;
	}
	\ForEach{$(psh\_node_m,psh\_node_c)\in post\_single\_hop\_node\_pairs$}{
	Add $psh\_node_m$ with its corresponding edges (from $aug_m$) to $aug'_m$\;
	Add $psh\_node_c$ with its corresponding edges (from $aug_c$) to $aug'_c$\;
	}	
	$aug'_m \gets hsu(aug'_m, \{order\_edge\})$\;
	$aug'_c \gets hsu(aug'_c, \{order\_edge\})$\;
	$(aug'_m \rightarrow aug'_c) \gets$ $aug'_m$ and $aug'_c$ connected with \lstinline[basicstyle=\ttfamily]|transform|-edges between nodes in $aug'_m$ and $aug'_c$ based on the mapping in $minMap$\;
\end{algorithm}

The algorithm receives the misuse AUG $aug_m$ and the fix AUG $aug_c$. 
It then produces a bipartite graph (cf. \autoref{alg:change_rule} Lines 1--5) by first equalizing the node cardinalities of both graphs by adding empty $\epsilon$-nodes (Lines 1--3). 
This bipartite graph consists of two partitions representing the nodes of $aug_m$ and $aug_c$, respectively. 
Then, all nodes between the two partitions are connected with edges labeled by the $cost$-function (Lines 4--5). 
The $cost$-function computes the number of edits (i.e., re-labeling and adding/deleting in- and outgoing edges) needed to transform one node into the respective other node. 
Afterward, we use the Kuhn-Munkres algorithm~\cite{Munkres1957} to find a one-to-one mapping with minimal overall costs, and construct the sub-AUGs $aug'_m$ and $aug'_c$ with only changed nodes---namely nodes for which $costs>0$ (cf. Lines 6--7). 

These steps of our algorithm have not been altered concerning our previous work~\cite{Nielebock2021b}. 
Still, in contrast to our initial idea of correction rules~\cite{Nielebock2020}, we do not require a user to manually add the misused API for filtering nodes in the AUG. 
This simplifies the automated inference of change rules. 
Nevertheless, we designed \tool{} to be easily extendable with such a feature in case it can further improve a rule's quality. 

Regarding the three challenges we identified, we further revised the algorithm as follows:
\begin{description}
	\item[Challenge 1:] To handle missing data nodes, we execute a post-hoc adaptation of the change rule. 
	First, we add all neighboring nodes of a changed action node that are connected via a data-flow edge. 
	So, we add all directly incoming parameters and outgoing result objects. 
	We denote this as \emph{single-hop addition} (cf. \autoref{alg:change_rule} Lines 8--12). 
	Moreover, we noticed some cases in which it is essential to know how these single-hop-added nodes are subsequently used in the AUG. 
	Thus, second, we also add for each single-hop-added node all neighboring nodes connected via an outgoing data-flow edge (cf. Lines 13--20).

	\item[Challenge 2:] To handle missing action nodes related to \lstinline[basicstyle=\ttfamily]|finally|-edges, we reuse the single-hop-addition to add all nodes connected to such an edge (cf. Lines 8--12).

	\item[Challenge 3:] To reduce the number of \lstinline[basicstyle=\ttfamily]|order|-edges, we apply a reduction step using \citeauthor{Hsu1975}'s algorithm~\cite{Hsu1975} to compute a minimal equivalent graph (MEG) as implemented in the JGraphT library.\footnote{\url{https://jgrapht.org/javadoc-1.5.0/org.jgrapht.core/org/jgrapht/alg/TransitiveReduction.html}}
	This algorithm minimizes the number of edges while assuring the graph's reachability and that the new graph remains a sub-graph of the original one.
	Note that JGraphT applies only the first part of \citeauthor{Hsu1975}'s algorithm, computing only the MEG for acyclic graphs. 
	Since a formal proof that AUGs are acyclic is missing, we cannot directly guarantee the applicability of this algorithm. 
	Thus, we only employ the \lstinline[basicstyle=\ttfamily]|order|-edge reduction (cf. Lines 21--22) if the AUG is acyclic. In our experiments, we did not observe a cyclic AUG. 
\end{description}
Finally, we add the \lstinline[basicstyle=\ttfamily]|transform| edges based on our minimal mapping $minMap$ to obtain the final change rule $aug'_m\rightarrow aug'_c$ (Line 23).

\subsection{Applicability Check}\label{ssec:applicability}

\myParNoSpace{Applicability Check ({\footnotesize\circled{E}})} 
In our previous work~\cite{Nielebock2021b}, we experimented with API misuse detection based on a set of correction rules that we found to be applicable to known correct API usages and API misuses. 
Particularly, we denoted a correction rule as applicable if the mean distance value between its misuse part is closer to the API misuses than to the fixed API usages, and its correct part is closer to the fixed API usages than to the API misuses. 
However, this strategy has some drawbacks. 
First, we need a valid set of known correct API usages and API misuses to compute the distances, which requires essential manual effort for preparation---similar to the efforts of creating API misuse benchmarks~\cite{Amann2016,Kang2021}. 
Second, even though we have a sufficiently large dataset, many rules are only applicable to one specific misuse of a single library, making them infeasible for detecting API misuses in a broader context. 
Finally, the elicited dataset may not contain any example of the particular API relevant for detecting a misuse, and thus we would discard that correction rule due to the missing applicability to the dataset.

We have relaxed the applicability check to only consider the API usage and misuse to be checked using our novel change rules. 
Particularly, we check whether the misuse part of a change rule is similar to the API usage under test. 
Formally, we check with a change rule $aug_{rm}\rightarrow aug_{rc}$ and an API usage to be tested ($aug_x$) whether
\begin{equation}\label{eq:applic}
	dist(aug_{rm},aug_{x})<threshold
\end{equation}
holds. 
The hyperparameter $threshold \in [0,1]$ is user-defined, for instance, based on the results of our evaluation that we report in \autoref{sec:evaluation}. 

\subsection{Graph Distance-Based API Misuse Detection}
\label{ssec:dist}

In the following, we describe how we compute different types of distances between API misuses and our change rules.
Then, we specify our actual API misuse detection based on these distances.

\myPar{Distance Computation {\footnotesize\circled{F}}}
From our previous experiments~\cite{Nielebock2021b}, we derived hypotheses for improving distance functions to achieve a better API misuse detection. 
In our current technique, we rely on Exas vectors, since these and their distance values are easily computable---scaling better for frequent distance computations as required for the use case of misuse detection. 
Next, we introduce a formal description of Exas vectors as well as a modified computation of the normalization based on the original work by \citet{Nguyen2009}. 
Afterward, we introduce our different extensions as well as their rationales to improve the misuse detection.

Formally, we denote $vec_A$ and $vec_B$ as the Exas vectors representing the AUGs $aug_A$ and $aug_B$, respectively. 
These vectors contain the frequencies of \emph{n-paths} and \emph{(p,q)-nodes} present in an AUG. 
The \emph{n}-paths features describe all paths up to length \emph{n} in an AUG, while \emph{(p,q)}-nodes describe all nodes from the AUG having \emph{p} incoming and \emph{q} outgoing edges. 
We identify these features based on their labels. 
For instance, in \autoref{fig:aug}, \lstinline[basicstyle=\ttfamily]|Object.<init>-1-6| would be one \emph{(p,q)}-node, while \lstinline[basicstyle=\ttfamily]|[Foo,| \lstinline[basicstyle=\ttfamily]|para, Object.<init>]| would be an \emph{n}-path of length two. 
Since \citeauthor{Nguyen2009}~\cite{Nguyen2009} achieved the best results by setting \emph{n} to four, we also compute the Exas vectors \emph{n}-paths up to this length. 
Moreover, we ignore \emph{n}-paths of length one (i.e., single nodes), since these are already represented by the respective \emph{(p,q)}-node.

\citeauthor{Nguyen2009}~\cite{Nguyen2009} proved that the L1-norm of the difference between two Exas vectors is related to the GED. 
Particularly, if two graphs have a certain GED, the L1-norm of their vector difference has an upper bound dependent on the GED. 
Note that \emph{the opposite does not hold}. 
This means that, to some extent, the length of the difference between two Exas vectors can be used to estimate the distance between the graphs they represent. 
Previously~\cite{Nielebock2021b}, we leveraged this property for the distance computation and proposed two distances: 
first, the L1-norm as suggested by \citeauthor{Nguyen2009} and, second, the cosine similarity. 
We then normalized these distances to obtain a relative distance in the interval $[0,1]$ to make different Exas vector distances comparable. 
Since these distances were not successful in detecting API misuses, we subsequently discuss a revised version.

Exas vectors only contain those features present in the respective AUG. 
Consequently, we need to equalize their feature sets to compute a valid difference. 
We do this by defining $\tilde{vec_A}$ and $\tilde{vec_B}$ as the \emph{sub-vectors} of $vec_A$ and $vec_B$, where the feature set is the cut of both original feature sets. 
To include the ratio between the number of matched and non-matched features, as well as the similarity of the frequencies of the matched features, we compute $feature_{dist}$ as well as $featureCount_{dist}$. 
The function $feature_{dist}$ represents how many features from one vector match with the features from the respective other vector. 
This way, we can quantify to what degree a vector is a sub-vector of the other one:
\begin{equation}\label{eq:feature_dist}
	feature_{dist}(vec_A,vec_B) = 1- max\left(\frac{len(\tilde{vec_A})}{len(vec_B)},\frac{len(\tilde{vec_B})}{len(vec_A)}\right)
\end{equation}
with $len$ being a function computing the length of a vector (i.e., number of features). 
Note that, by definition, $len(\tilde{vec_A})=len(\tilde{vec_B})$ holds.

To compute $featureCount_{dist}$, we use two distance functions, namely the L1-norm and the cosine similarity, and normalize them to ensure a relative distance computation.
We compute both distances as follows:
\begin{equation}\label{eq:featurecount_dist_l1}
	featureCount_{dist-L1}(vec_A,vec_B) = \left|\left|\frac{\tilde{vec_A}-\tilde{vec_B}}{max(1, maxVal(\tilde{vec_A}-\tilde{vec_B}))} \right|\right|_{1}
\end{equation}
where $maxVal$ computes the maximum absolute value within a vector, and
\begin{equation}\label{eq:featurecount_dist_cosine}
	featureCount_{dist-Cosine}(vec_A,vec_B) = 1-\frac{\langle\tilde{vec_A}, \tilde{vec_B}\rangle}{||\tilde{vec_A}||_2||\tilde{vec_B}||_2}
\end{equation}
where $\langle\cdot,\cdot\rangle$ denotes the scalar product. 
Our final $dist$ function is a combination of $feature_{dist}$ and $featureCount_{dist}$:
\begin{equation}
	\label{eq:dist}
	\begin{split}
dist_{ExasVector}(aug_A,aug_B) &= \lambda \cdot feature_{dist}(vec_A,vec_B)\\ &+ (1-\lambda) \cdot  featureCount_{dist}(vec_A,vec_B)\\ &\lambda \in [0,1)
	\end{split}
\end{equation}
In our evaluation, we used $0.5$ for the scaling factor $\lambda$. 
We compute $dist_{ExasVectorL1Norm}$ and $dist_{ExasVectorCosine}$ by applying the respective $featureCount_{dist-L1}$ and $featureCount_{dist-Cosine}$ function, respectively.
Please note that these equations are similar, but not identical, to our previous work~\cite{Nielebock2021b}---we refined them to be symmetric.

Based on our previous hypotheses~\cite{Nielebock2021b}, we implemented the following variants of the distance functions to research whether they can improve the API misuse detection:
\begin{description}
	\item[Indicator Over Frequency:] 
	When normalizing Exas vectors, we noticed that vectors containing single highly frequent features hide the relative feature frequency of less frequent features. 
	So, some features became less important for the final distance computation. 
	To mitigate this effect, we propose to use indicators that denote whether a certain feature is present (i.e., 1) or not (i.e., 0), instead of using absolute frequencies. 
	The computation of the distance values remains the same.
	Note that this adaptation may hide that certain features must be present a certain number of times.
	We denote indicator vectors with the prefix \texttt{Indicator-} in the distance function name.

	\item[API-Specific Features:] 
	We intend to use the computed distance values for API misuse detection. 
	However, some vector distances were mainly influenced by features, and thus AUG nodes, that do not refer to any external API. 
	Therefore, we propose to analyze only those features that contain information about an API. 
	For this purpose, we introduce a function $api$ that takes as input a node from an AUG and returns the referred API class if available (cf. step {\footnotesize\circled{C}}). 
	For each feature that contains at least one label of an AUG node with non-empty API-type information, we keep that feature in the Exas vector. 
	Then, we compute the remaining distance as before. 
	We denote distance functions building on API-specific features with the prefix \texttt{API-}.

	\item[API-Specific Vector Splitting] 
	Some API usages intertwine different APIs in one method, which would hamper the detection of a similar API usage since non-related APIs are present. 
	We suggest a splitting mechanism that divides an AUG based on its contained API types into several sub-AUGs. 
	Particularly, we group nodes and their respective edges from the original AUG based on the package name of their related API type, namely up to the first three entries of the package name (e.g., \lstinline[basicstyle=\ttfamily]|java.lang| for \lstinline[basicstyle=\ttfamily]|java.lang.Object|). This package name is also used as label of the respective sub-AUG. 
	If we have only a class name or no API information at all, we group those AUG elements in a special miscellaneous sub-AUG. 
	Then, we compute each sub-distance between the sub-AUGs with the same labeled group.
	The overall distance is the average of all non-distinct (i.e., $dist\neq1$) sub-distances. 
	We denote this distance function with the infix \texttt{-Split-}.

	\item[Combinations] 
	We combine all extensions to analyze whether this improves the overall performance of our technique. 
	This provides 16 different distance computations, namely two norms (i.e., L1 and Cosine) for two count types (i.e., indicator vs. frequency) for two feature types (i.e., API-specific vs. all) for two splitting types (i.e., splitting vs. no splitting). 
	However, some of the combinations are redundant. 
	Particularly, using an \texttt{Indicator} function in \autoref{eq:dist} causes the $featureCount_{dist}$-function to be either $1$ or $0$, since this computes the differences of the sub-vectors that contain only features present in both AUGs, which are all $1$. 
	So, the actual norms (i.e., L1Norm and Cosine) are obsolete. 
	For this reason, we sum up \texttt{Indicator} functions by naming them without the actual norm; reducing the eight cases to four. 
	So, we employ 12 different combinations during our evaluation.
\end{description}
We analyze and compare the impact of these adaptations through our evaluation in \autoref{sec:evaluation}.

\myPar{API Misuse Detection (\footnotesize\circled{G})}
If a change rule is applicable, we use \autoref{eq:misuse-detect} to decide whether the checked API usage is a potential misuse.
To validate this decision, we also need to define what denotes a correct result. 
For instance, assume we have a known API misuse and two applicable rules $rule_1$ and $rule_2$, from which $rule_1$ detects the misuse and $rule_2$ does not. 
Then, $rule_1$ produces a true positive ($tp$) result while $rule2$ produces a false negative ($fn$) result. 
Differently, assume a correct API usage for which the same rules $rule_1$ and $rule_2$ are applicable. 
For this usage, $rule_1$ detects the misuse, while $rule_2$ decides the API usage is no misuse. 
We say that $rule_1$ produces a false positive ($fp$) result while $rule_2$ produces a true negative ($tn$) result. 

For each rule, we can compute the number of $tp$s, $tn$s, $fp$s, and $fn$s based on a set of known misuses and correct usages---and subsequently precision and recall. 
Precision measures the proportion of $tp$ decisions of a rule to all positive decisions. 
So, it provides an estimation of how trustful the detected API misuses are (i.e., whether they avoid \enquote{false alarms}).  
Recall measures the proportion of detected misuses from the set of known misuses. 
Please note that our main goal is to increase the precision, while we expect the recall to be low. 
This is reasonable, since change rules depict a very specific API misuse fix, and thus hardly generalize.
\section{Evaluation\label{sec:evaluation}}

We now present our evaluation. 
To enable replications and reproductions, we publish our complete implementation including \tool{}, our graph vectorization, distance computation, previous rule analysis, and evaluation results in a persistent, publicly available Zenodo repository.\textsuperscript{\ref{fn:replication}}

\subsection{Datasets and Experimental Setting}
\label{ssec:datasets}

\myPar{Datasets}
We evaluate our API misuse detection on three independent and publicly available datasets of real API misuses from the open-source domain.
\begin{description}
	\item[MUBench] is a dataset developed by \citet{Amann2016} using manually evaluated fixes of API misuses.\footnote{\url{https://github.com/stg-tud/MUBench}, accessed on January \nth{27}, 2022} 
	It comprises fixing commits for 280 API misuses, together with their repository URIs. 
	We only used those 116 entries that we used in our previous work~\cite{Nielebock2021b}. 
	Moreover, for two entries, we could not retrieve the respective commits anymore, reducing the number to 114 entries. 
	Each API-misuse entry lists the respective git repository, fixing commit hash, as well as the declaration of the misuse-containing method and the respective source-file path.

	\item[AU500] is a dataset of 500 manually assessed API usages by \citet{Kang2021}.\footnote{\url{https://github.com/ALP-active-miner/ALP}, accessed on January \nth{27}, 2022} 
	It comprises 385 correct API usages and 115 API misuses. 
	AU500 lists the repository, commit hash, source file, method name, and line number of the usage together with a manual assessment of the ground truth (i.e., correct usage or misuse). 
	Please note that when we constructed the AUGs for AU500 using the implementation of \citet{Amann2019a}, we could only generate these for 493 entries (379 correct usages and 114 misuses), identically to our previous work~\cite{Nielebock2021b}.

	\item[AndroidCompass] is our own dataset~\cite{Nielebock2021} of 80,324 changed Android compatibility checks together with their respective repository, commit hashes, source-file paths, and source line.\footnote{\url{https://doi.org/10.5281/zenodo.4428340} accessed on January \nth{27}, 2022} 
	These compatibility checks represent a common pattern in Android apps. 
	Particularly, they protect apps from calling APIs from Android versions that are not present in the version installed on the executing hardware. 
	So, the edits to the \lstinline[basicstyle=\ttfamily]|if|-statements can be interpreted as API misuse fixes. 
	In our experiments, we could produce 24,610 non-empty change rules for the 80,324 entries.
\end{description}
We used these datasets to conduct our experiments on different ground truths, which are not particularly tailored to our misuse detector. This is important for a realistic evaluation in the context of misuse detection as seen in the program repair domain~\cite{Durieux2019ProgramRepair}.

\myPar{Experimental Setting}
To evaluate our API misuse detection, we measure the impact of the 12 different combinations of distance functions (cf. \autoref{ssec:dist}) as well as of different $threshold$ values for the applicability check (cf. \autoref{ssec:applicability}). 
We measure the precision and recall for individual change rules using the number of $tp$, $fp$, $tn$, and $fn$ results as described in \autoref{ssec:dist}. 
For the overall performance, we build on the distributions as well as means among all rules. 
When comparing different distance functions (cf. \autoref{ssec:dist}) and $threshold$ values for the applicability check (cf. \autoref{ssec:applicability}), we compare those distributions using statistical tests.
When decreasing the threshold for the applicability check, we also decrease the number of applicable rules, and thus the number of rules producing positive results. 
This impacts the overall precision, since we do not achieve any positive results for some thresholds (i.e., $\#tp+\#fp=0$). 
So, we cannot compute the mean precision among all rules, but only among those that produce positive results. 
We refer to this precision as \emph{relative precision}. 
However, we can hardly compare those values, since the actual number of applicable rules may differ between the different distance functions. 
To mitigate this problem, we compute a \emph{conservative precision} that sets all precision values to $0$ if $\#tp+\#fp=0$. 
Note that this is conservative because we consider rules producing neither $tp$s nor $fp$s as equal with those rules producing only $fp$s.

In our first two experiments, we investigate the best threshold values regarding \autoref{eq:applic} by conducting a grid-search-based method; spanning the threshold from $<1.1$ to $<0.1$ in $0.1$-steps. 
Based on the results, in the third experiment, we use a threshold of $0.4$ and $0.2$ to reduce the computational effort. 
In detail, we report on the following three experiments:
\begin{description}
	\item[Experiment 1: MUBench on MUBench.] In the first experiment, we derive change rules from each entry in the MUBench dataset and apply these to the misuse and fixed versions of all other entries in MUBench. 
	Note that we do not apply the rule on its own misuse and fix, since this would bias the $tp$ and $tn$. 
	Nevertheless, we found that MUBench has a significant proportion of similar misuses from the Joda-time project~\cite{Nielebock2021a}.
	Thus, we expect that the results may be positively influenced by having similar misuses from the same project. 
	These results should be representative of reusing fixed API misuses for project-internal API misuse detection. 
	Since MUBench consists of misuses from multiple projects, we can also partially observe the cross-project applicability of our techniques.

	\item[Experiment 2: MUBench on AU500.] In the second experiment, we apply the previously obtained change rules from MUBench on the AU500 dataset. 
	We check whether the rules can correctly detect the API misuses and do not report correct usages as misuses. 
	Since MUBench and AU500 do not share any projects, we consider the results  representative for reusing fixed API misuses in a cross-project manner.

	\item[Experiment 3: AndroidCompass on AndroidCompass.] In the last experiment, we use the AndroidCompass dataset to obtain change rules. 
	We assume that the version before an updated compatibility check is the misuse, while the updated version is the respective fix. 
	Next to the rules, we also compute the misuse and fix AUGs for each entry. 
	We validate the API misuse detection by conducting a ten-fold cross-validation. 
	Particularly, we split all entries into ten subsets (i.e., buckets) and use the change rules obtained from nine buckets to detect API misuses in the remaining bucket. 
	We sample the buckets so that entries from one repository are contained in exactly one bucket. 
	To achieve this, we define a maximum size of entries in a bucket (i.e., $\frac{\#entries}{\#buckets}$). 
	Then, we sort the repositories according to their number of entries in AndroidCompass. 
	In a round-robin-like method, we iterate over the repositories and assign all entries of one repository to the respective buckets (i.e., all entries of the largest repository go into bucket \#1, all entries of the second-largest repository go into bucket \#2, and so on) as long as the maximum size is not exceeded. 
	
	During initial experimenting with the generated AUGs, we observed a significantly increased run time for computing distances for larger AUGs and change rules (in terms of the number of nodes). 
	So, we decided to limit the analysis to AUGs and rules with fewer than 100 nodes. 
	We selected this number through an interactive testing method rather than a systematic analysis. 
	Nevertheless, it turned out to be a good compromise between the amount data to analyze and the computation time. 
	Also, rules and AUGs with 100 nodes or more usually represent a very specific and complex change, which is unlikely to be reusable in other projects. 
	Due to this limit, we consider 16,094 entries from 931 different repositories for our analysis, for which we display descriptive statistics in \autoref{tab:descrip_buckets}.
	
	We compute the performance of each run by calculating the mean of the performance metrics of all rules from the respective nine buckets. 
	For the overall performance, we present the distribution of these mean values as well as the mean over all ten runs.
	Since AndroidCompass provides misuse fixes for similar misuses, namely proper handling of Android compatibility checks, this experiment provides insights into the cross-project applicability of our techniques.
\end{description}
Before conducting our evaluation, we assessed for each dataset from which we obtain change rules (i.e., MUBench, AndroidCompass) the proportion of non-empty change rules that could be computed from all entries (cf. \autoref{ssec:cri_evaluation}).

\begin{table}
\caption{Overview of the data we use from AndroidCompass.}\label{tab:descrip_buckets}
\begin{tabular}{rrr}
	\hline
	\textbf{bucket} & \textbf{\#entries} & \textbf{\#repositories}\\
	\hline
	1 & 1,610 & 44\\
	2 & 1,610 & 52\\
	3 & 1,610 & 60\\
	4 & 1,610 & 66\\						
	5 & 1,609 & 75\\
	6 & 1,609 & 88\\
	7 & 1,609 & 102\\
	8 & 1,609 & 116\\
	9 & 1,609 & 138\\
	10 & 1,609 & 190\\\hline
	\textbf{all} & \textbf{16,094} & \textbf{931}\\\hline
\end{tabular}
\end{table}

\subsection{Change Rule Inference\label{ssec:cri_evaluation}}
Next, we present the results for constructing change rules from MUBench and AndroidCompass using \tool{}, particularly \autoref{alg:change_rule}.
Regarding MUBench, we had an initial set of 114 entries from which we constructed change rules. 
For that purpose, we required the respective AUGs of the API misuse and fixed version. 
We could construct those for 92 entries. 
\tool{} was able to produce 89 non-empty change rules (i.e., $\approx 78\,\%$ of all analyzed entries and $\approx 96\,\%$ of all generated AUGs).

Regarding AndroidCompass, we filtered the 80,324 entries to exclude:
\begin{itemize}
	\item 40 entries that we previously used for testing purposes;
	\item entries whose commits only add or delete source files, since the changes are too large; and
	\item entries for which we could not determine the respective version code of the compatibility check (i.e., the Android version against which the code is tested), since the missing type resolution hampers the AUG generation.
\end{itemize}
This filtering left 51,999 entries from which we could produce 38,347 misuse and 41,096 fix AUGs. 
The respective cut set contained 35,857 complete (i.e., misuse and fix AUG are present) entries. 
From these entries, \tool{} could construct 24,610 non-empty change rules ($\approx 47\,\%$ of all tested entries in the dataset and $\approx 68\,\%$ of all generated AUGs). 
Note that in 6,875 cases the rule generation produced an empty rule, while in the remaining 4,372 cases the rule generation was aborted.\footnote{We did not log the concrete exceptions, and thus cannot judge the exact reasons for this behavior.} 
Due to the size filtering (i.e., we discarded AUGs and rules with more than 100 nodes), the number of entries we used in the experiment further decreased to 16,094 change rules.

We observed that particularly for AndroidCompass the rule inference mainly depends on a successful AUG generation. 
Many change rules could not be created because we were not able to produce the respective AUGs. 
While we did not investigate the core reasons for this problem, we assume that this is likely due to parsing errors. 
The original AUG generation targets Java 8, and even though we updated the parser to handle Java 11 some newly introduced concepts may not be properly handled in the AUG generation. 
Nevertheless, a majority of change rules could be inferred if the AUGs could be generated, making \tool{} in principle applicable for API misuse detection.

\subsection{API Misuse Detection}
\label{ssec:eval_misuse_detect}

Now, we report and discuss the results of the three experiments we described in \autoref{ssec:datasets}.

\begin{figure}
	\begin{subfigure}{.49\textwidth}
		\caption{Relative Precision\label{fig:rel_precision_mubench}}
		\includegraphics[width=\textwidth, trim=58 58 70 100, clip]{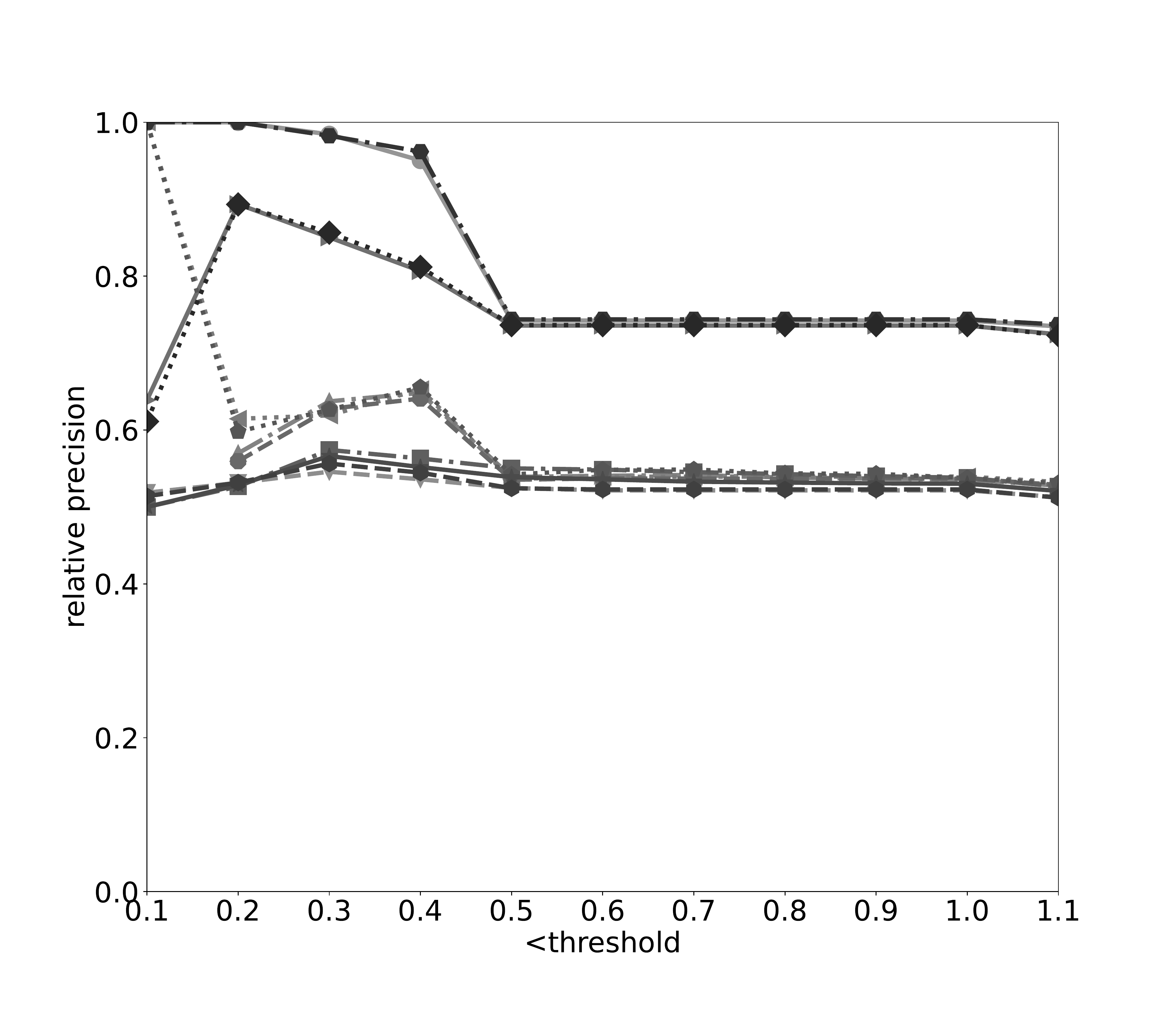}	
	\end{subfigure}	
	\begin{subfigure}{.49\textwidth}
		\caption{Conservative Precision\label{fig:cons_precision_mubench}}
		\includegraphics[width=\textwidth, trim=58 58 70 100, clip]{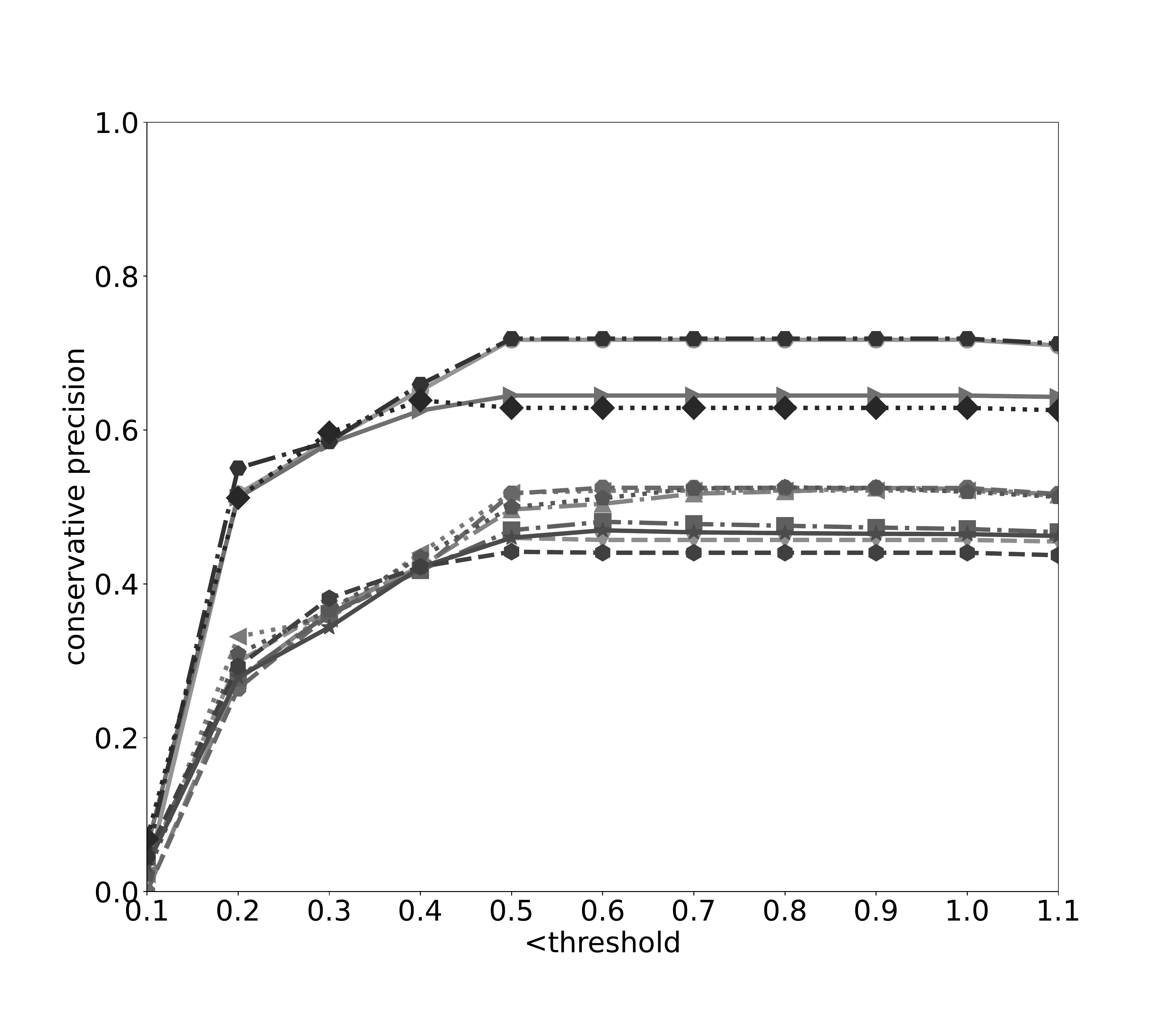}
	\end{subfigure}
	\begin{subfigure}{.49\textwidth}
		\caption{Recall\label{fig:recall_mubench}}
		\includegraphics[width=\textwidth, trim=58 58 70 100, clip]{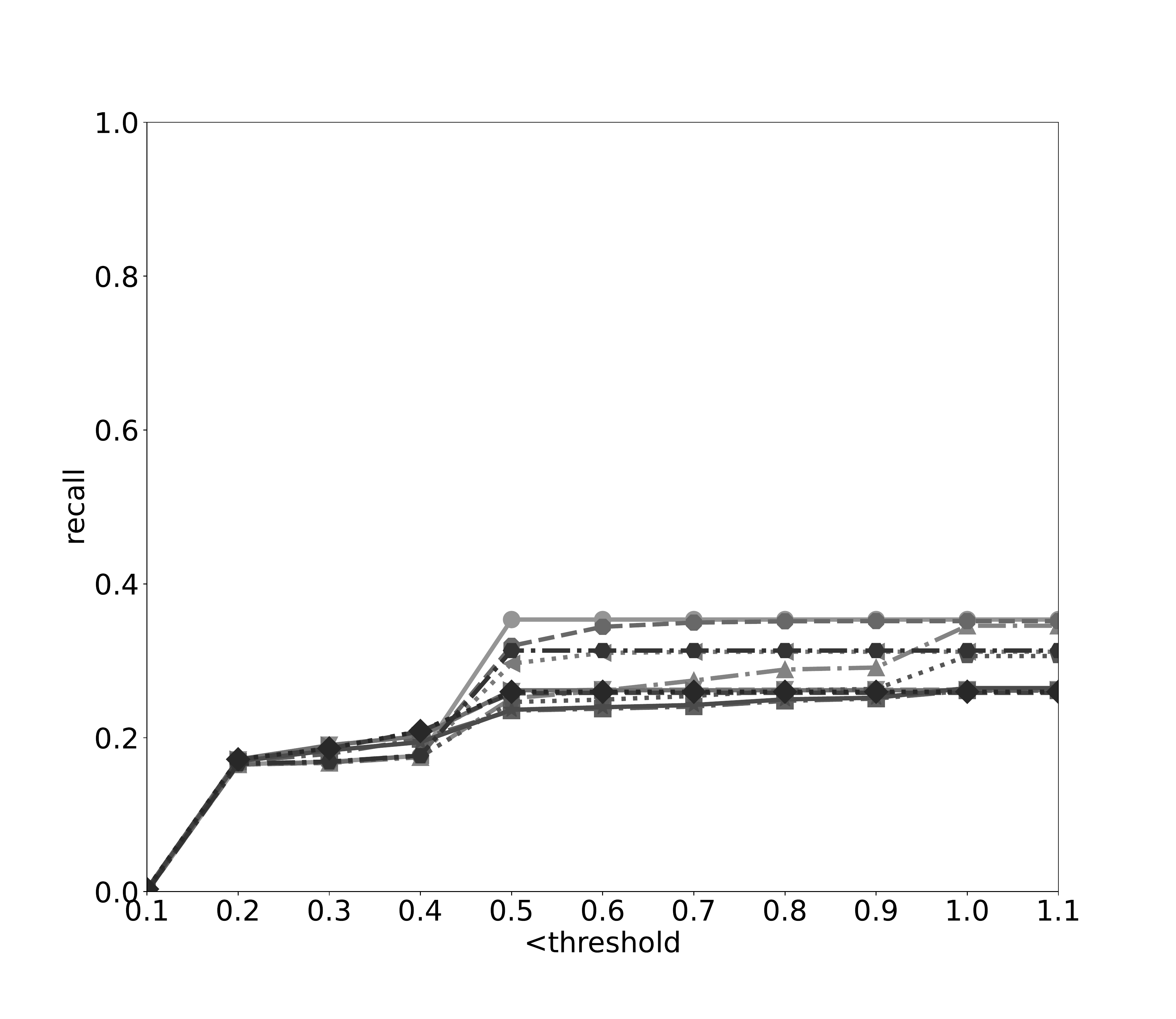}
	\end{subfigure}	
	\begin{subfigure}{\textwidth}
		\includegraphics[width=\textwidth, trim=70 140 70 50, clip]{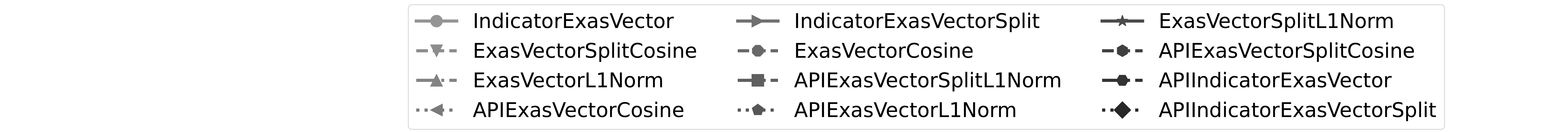}
	\end{subfigure}	
	\caption{Mean performance values on MUBench using different thresholds for the applicability check.}
	\label{fig:misuse_mubench}
\end{figure}

\begin{figure}
	\includegraphics[width=.7\textwidth, trim=60 60 100 100, clip]{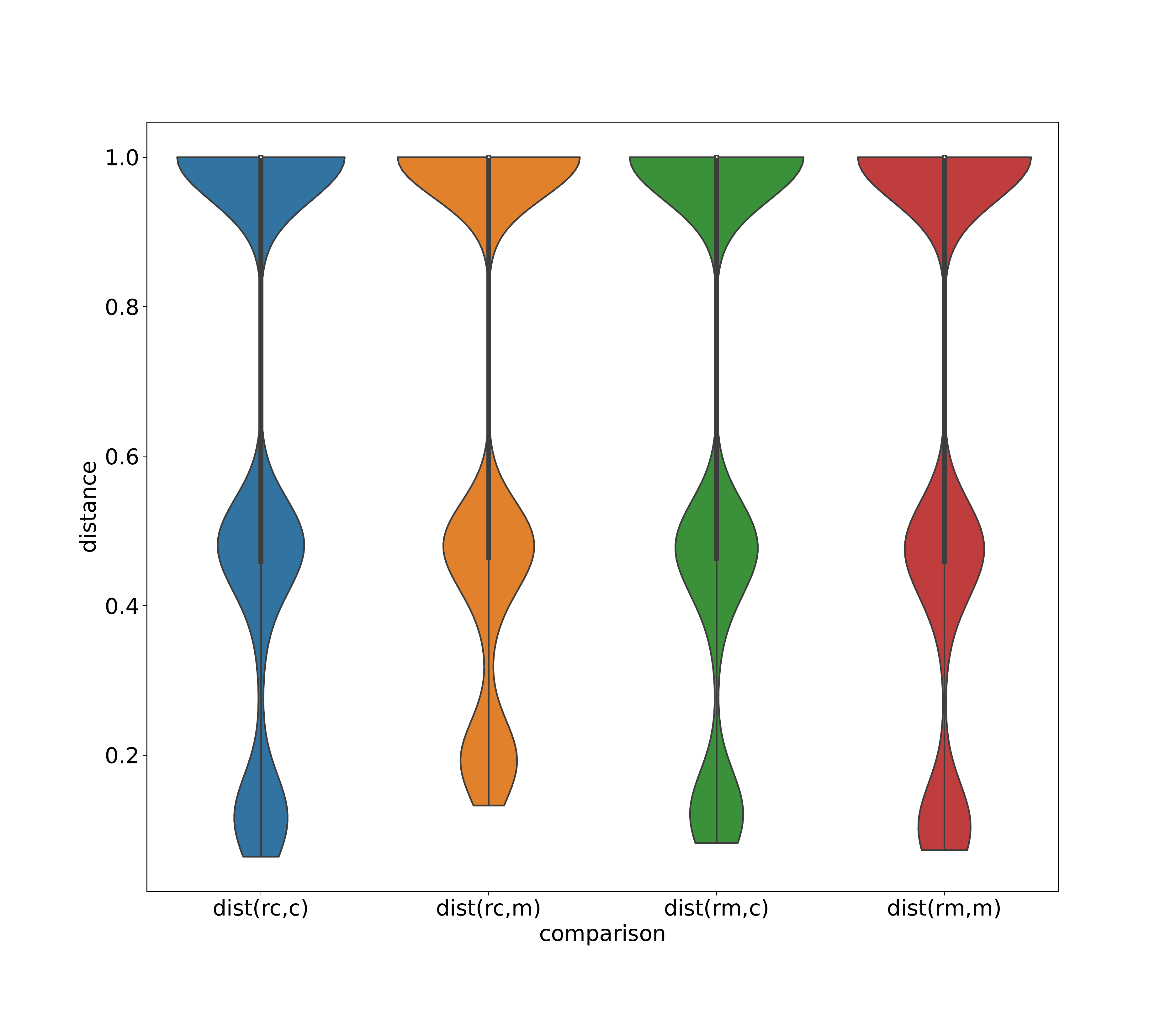}
	\caption{Violin plots of the distribution of distance values when comparing the parts of the change rule (i.e., $rm \rightarrow rc$) with a misuse (i.e., $m$) and a correct usage (i.e., $c$) using \texttt{IndicatorExasVector}.}
	\label{fig:violin_plot_dist}
\end{figure}

\myPar{Experiment 1: MUBench on MUBench}
We computed the performance values for each change rule as described in \autoref{ssec:datasets} and depict the mean values for the different thresholds in \autoref{fig:misuse_mubench}. 
In general, we observe that the mean values are more dynamic with thresholds of $<0.5$.
This is caused by most distance values falling into the range of $[0.5,1]$, while only a few have a value of $0.5$ or lower. 
For instance, we depict a violin plots of the different distance values computed with the \texttt{IndicatorExasVector} in \autoref{fig:violin_plot_dist}---all other distance functions result in similar distributions. 
We can see that more distance values are located near $1$ and $0.5$. 
So, when filtering out a major set of rules, the performance metrics are only influenced by the much smaller distance values located in the range $[0,0.5)$. 

More distances being around the values $1$ and $0.5$ can be explained with one property of \autoref{eq:dist}.
Since most API usages are different, they do not share any features (i.e., $feature_{dist}=1$), and thus no shared sub-vectors exist (i.e., $featureCount_{dist}=1$). 
However, assume that two vectors have only one matching feature. 
Then, the frequency of features is likely to be equal (i.e., $featureCount_{dist}\approx0$), while the proportion of matched features to all features in a vector is usually low (i.e., $feature_{dist}\approx1$). 
So, a likely distance value is $\lambda$, and because we defined $\lambda=0.5$, this value occurs more frequently.

When comparing relative (\autoref{fig:rel_precision_mubench}) and conservative precision (\autoref{fig:cons_precision_mubench}) to the recall (\autoref{fig:recall_mubench}), we can see that the mean precision is usually higher than the mean recall; which we expected. 
The recall is constantly decreasing with smaller thresholds, because more and more rules are not applicable, and thus do not detect any positive results. 
Similarly, the conservative precision decreases, since we assign a precision of zero to non-applicable rules.

We can see in \autoref{fig:rel_precision_mubench} that the two distance functions \texttt{In\-di\-ca\-tor\-Exas\-Vec\-tor} and \texttt{API\-In\-di\-ca\-tor\-Exas\-Vec\-tor} achieved a relative precision of more than $0.9$ (i.e., $\approx95\,\%$ and $\approx96.1\,\%$, respectively) with a threshold of $<0.4$. 
Moreover, their respective \texttt{-Split-} versions achieved the maximum of $\approx89.3\,\%$ at a threshold of $<0.2$. 
For the threshold $<0.1$, the distance functions \texttt{API\-Exas\-Vec\-tor\-L1Norm} and \texttt{API\-Exas\-Vec\-tor\-Co\-sine} achieved a mean precision of $100\,\%$. 
Considering the respective conservative precision (cf. \autoref{fig:cons_precision_mubench}), the distance functions \texttt{IndicatorExasVector} and \texttt{APIIndicatorExasVector} usually achieved the highest precision values (i.e., $\approx 71.7\,\%$ for both at the threshold $<0.5$) followed by the their respective \texttt{-Split-} functions (i.e.,  $\approx 64.4\,\%$ for \texttt{IndicatorSplitExasVector} at the threshold $<0.5$ and $\approx 63.8\,\%$ for \texttt{API\-In\-di\-ca\-tor\-Split\-Exas\-Vec\-tor} at the threshold $<0.4$) .

Based on this observation, we estimate the best performance for the distance functions in the threshold range $[0.1,0.4]$. 
Since the recall is constantly decreasing with smaller thresholds, we further checked the maximum threshold value of $<0.4$. 
Particularly, we analyzed whether the differences in the means for the conservative precision and recall at this threshold are statistically significant. 
We used the Wilcoxon signed-rank test ($\alpha=0.05$) to conduct pairwise comparisons, since this test does not enforce a normal distribution. 
To cope with zero differences in the ranks, we applied the \enquote{pratt}-option by scipy\footnote{\url{https://docs.scipy.org/doc/scipy/reference/generated/scipy.stats.wilcoxon.html}}. 
Because we have multiple comparisons, we applied the conservative Bonferroni correction on all statistical tests. 
We depict the results of our tests in \autoref{tab:s_tests_prec_mubench} and \autoref{tab:s_tests_rec_mubench}. 
Note that we conducted the comparisons only once, but depict the mirrored test results (i.e., distance function A compared to B is identical to B compared with A).

\begin{table}
	\caption{Statistical test results for MUBench on MUBench regarding the pairwise comparison of the conservative precision at threshold $<0.4$ using the Wilcoxon signed-rank test ($\alpha=0.05$) with a Bonferroni correction. \cmark\ denotes the significant results, while \xmark\ denotes the non-significant results. The $>$ and $<$ characters denote whether the mean conservative precision of the function on the left-hand side is $<$ or $>$ compared to the function on the top. Numbers after significant results represent the effect size measured with Cliff's $\delta$.}\label{tab:s_tests_prec_mubench}
\begin{adjustbox}{max width=\linewidth}	
\begin{tabular}{l|cccccccccccc}
	\toprule
	{} & \rotatebox{90}{APIExasVectorCosine} & \rotatebox{90}{APIExasVectorL1Norm} & \rotatebox{90}{APIExasVectorSplitCosine} & \rotatebox{90}{APIExasVectorSplitL1Norm} & \rotatebox{90}{APIIndicatorExasVector} & \rotatebox{90}{APIIndicatorExasVectorSplit} & \rotatebox{90}{ExasVectorCosine} & \rotatebox{90}{ExasVectorL1Norm} & \rotatebox{90}{ExasVectorSplitCosine} & \rotatebox{90}{ExasVectorSplitL1Norm} & \rotatebox{90}{IndicatorExasVector} & \rotatebox{90}{IndicatorExasVectorSplit} \\
	\midrule
	APIExasVectorCosine         &                   - &              \xmark &                   \xmark &                   \xmark &         < \cmark -0.32 &               < \cmark -0.3\phantom{0} &           \xmark &           \xmark &                \xmark &                \xmark &      < \cmark -0.31 &           < \cmark -0.28 \\
	APIExasVectorL1Norm         &              \xmark &                   - &                   \xmark &                   \xmark &         < \cmark -0.32 &              < \cmark -0.31 &           \xmark &           \xmark &                \xmark &                \xmark &      < \cmark -0.31 &           < \cmark -0.29 \\
	APIExasVectorSplitCosine    &              \xmark &              \xmark &                        - &                   \xmark &         < \cmark -0.37 &              < \cmark -0.36 &           \xmark &           \xmark &                \xmark &                \xmark &      < \cmark -0.36 &           < \cmark -0.34 \\
	APIExasVectorSplitL1Norm    &              \xmark &              \xmark &                   \xmark &                        - &         < \cmark -0.38 &              < \cmark -0.36 &           \xmark &           \xmark &                \xmark &                \xmark &      < \cmark -0.36 &           < \cmark -0.33 \\
	APIIndicatorExasVector      &       > \cmark 0.32 &       > \cmark 0.32 &            > \cmark 0.37 &            > \cmark 0.38 &                      - &                      \xmark &    > \cmark 0.34 &    > \cmark 0.33 &         > \cmark 0.37 &         > \cmark 0.38 &              \xmark &                   \xmark \\
	APIIndicatorExasVectorSplit &        > \cmark 0.3\phantom{0} &       > \cmark 0.31 &            > \cmark 0.36 &            > \cmark 0.36 &                 \xmark &                           - &    > \cmark 0.33 &    > \cmark 0.32 &         > \cmark 0.36 &         > \cmark 0.35 &              \xmark &                   \xmark \\
	ExasVectorCosine            &              \xmark &              \xmark &                   \xmark &                   \xmark &         < \cmark -0.34 &              < \cmark -0.33 &                - &           \xmark &                \xmark &                \xmark &      < \cmark -0.33 &           < \cmark -0.31 \\
	ExasVectorL1Norm            &              \xmark &              \xmark &                   \xmark &                   \xmark &         < \cmark -0.33 &              < \cmark -0.32 &           \xmark &                - &                \xmark &                \xmark &      < \cmark -0.32 &            < \cmark -0.3\phantom{0} \\
	ExasVectorSplitCosine       &              \xmark &              \xmark &                   \xmark &                   \xmark &         < \cmark -0.37 &              < \cmark -0.36 &           \xmark &           \xmark &                     - &                \xmark &      < \cmark -0.36 &           < \cmark -0.34 \\
	ExasVectorSplitL1Norm       &              \xmark &              \xmark &                   \xmark &                   \xmark &         < \cmark -0.38 &              < \cmark -0.35 &           \xmark &           \xmark &                \xmark &                     - &      < \cmark -0.37 &           < \cmark -0.33 \\
	IndicatorExasVector         &       > \cmark 0.31 &       > \cmark 0.31 &            > \cmark 0.36 &            > \cmark 0.36 &                 \xmark &                      \xmark &    > \cmark 0.33 &    > \cmark 0.32 &         > \cmark 0.36 &         > \cmark 0.37 &                   - &                   \xmark \\
	IndicatorExasVectorSplit    &       > \cmark 0.28 &       > \cmark 0.29 &            > \cmark 0.34 &            > \cmark 0.33 &                 \xmark &                      \xmark &    > \cmark 0.31 &     > \cmark 0.3\phantom{0} &         > \cmark 0.34 &         > \cmark 0.33 &              \xmark &                        - \\
	\bottomrule
\end{tabular}
\end{adjustbox}
\end{table}

\begin{table}
	\caption{Statistical test results for MUBench on MUBench regarding the pairwise comparison of the recall at threshold $<0.4$ using the Wilcoxon signed-rank test ($\alpha=0.05$) with a Bonferroni correction. \cmark\ denotes the significant results, while \xmark\ denotes the non-significant results. The $>$ and $<$ character denotes whether the mean recall of the function on the left hand side is $<$ or $>$ compared to the function on the top. Numbers after significant results represent the effect size measured with Cliff's $\delta$.}\label{tab:s_tests_rec_mubench}
	\begin{adjustbox}{max width=\linewidth}	
		\begin{tabular}{l|cccccccccccc}
			\toprule
			{} & \rotatebox{90}{APIExasVectorCosine} & \rotatebox{90}{APIExasVectorL1Norm} & \rotatebox{90}{APIExasVectorSplitCosine} & \rotatebox{90}{APIExasVectorSplitL1Norm} & \rotatebox{90}{APIIndicatorExasVector} & \rotatebox{90}{APIIndicatorExasVectorSplit} & \rotatebox{90}{ExasVectorCosine} & \rotatebox{90}{ExasVectorL1Norm} & \rotatebox{90}{ExasVectorSplitCosine} & \rotatebox{90}{ExasVectorSplitL1Norm} & \rotatebox{90}{IndicatorExasVector} & \rotatebox{90}{IndicatorExasVectorSplit} \\
			\midrule
			APIExasVectorCosine         &                   - &              \xmark &           < \cmark -0.12 &                   \xmark &                 \xmark &              < \cmark -0.12 &           \xmark &           \xmark &        < \cmark -0.11 &        < \cmark -0.08 &              \xmark &           < \cmark -0.11 \\
			APIExasVectorL1Norm         &              \xmark &                   - &           < \cmark -0.12 &                   \xmark &                 \xmark &              < \cmark -0.13 &           \xmark &           \xmark &        < \cmark -0.12 &         < \cmark -0.1 &              \xmark &           < \cmark -0.12 \\
			APIExasVectorSplitCosine    &       > \cmark 0.12 &       > \cmark 0.12 &                        - &            > \cmark 0.03 &                 \xmark &                      \xmark &    > \cmark 0.12 &    > \cmark 0.13 &                \xmark &                \xmark &       > \cmark 0.11 &                   \xmark \\
			APIExasVectorSplitL1Norm    &              \xmark &              \xmark &           < \cmark -0.03 &                        - &                 \xmark &              < \cmark -0.04 &    > \cmark 0.09 &     > \cmark 0.1 &                \xmark &                \xmark &              \xmark &                   \xmark \\
			APIIndicatorExasVector      &              \xmark &              \xmark &                   \xmark &                   \xmark &                      - &              < \cmark -0.12 &           \xmark &           \xmark &        < \cmark -0.11 &                \xmark &              \xmark &           < \cmark -0.11 \\
			APIIndicatorExasVectorSplit &       > \cmark 0.12 &       > \cmark 0.13 &                   \xmark &            > \cmark 0.04 &          > \cmark 0.12 &                           - &    > \cmark 0.13 &    > \cmark 0.14 &                \xmark &                \xmark &       > \cmark 0.12 &                   \xmark \\
			ExasVectorCosine            &              \xmark &              \xmark &           < \cmark -0.12 &           < \cmark -0.09 &                 \xmark &              < \cmark -0.13 &                - &           \xmark &        < \cmark -0.12 &        < \cmark -0.09 &              \xmark &           < \cmark -0.12 \\
			ExasVectorL1Norm            &              \xmark &              \xmark &           < \cmark -0.13 &            < \cmark -0.1 &                 \xmark &              < \cmark -0.14 &           \xmark &                - &        < \cmark -0.13 &         < \cmark -0.1 &              \xmark &           < \cmark -0.13 \\
			ExasVectorSplitCosine       &       > \cmark 0.11 &       > \cmark 0.12 &                   \xmark &                   \xmark &          > \cmark 0.11 &                      \xmark &    > \cmark 0.12 &    > \cmark 0.13 &                     - &                \xmark &       > \cmark 0.11 &                   \xmark \\
			ExasVectorSplitL1Norm       &       > \cmark 0.08 &        > \cmark 0.1 &                   \xmark &                   \xmark &                 \xmark &                      \xmark &    > \cmark 0.09 &     > \cmark 0.1 &                \xmark &                     - &       > \cmark 0.08 &                   \xmark \\
			IndicatorExasVector         &              \xmark &              \xmark &           < \cmark -0.11 &                   \xmark &                 \xmark &              < \cmark -0.12 &           \xmark &           \xmark &        < \cmark -0.11 &        < \cmark -0.08 &                   - &           < \cmark -0.11 \\
			IndicatorExasVectorSplit    &       > \cmark 0.11 &       > \cmark 0.12 &                   \xmark &                   \xmark &          > \cmark 0.11 &                      \xmark &    > \cmark 0.12 &    > \cmark 0.13 &                \xmark &                \xmark &       > \cmark 0.11 &                        - \\
			\bottomrule
		\end{tabular}
	\end{adjustbox}
\end{table}

Regarding the conservative precision, we can observe that the functions \texttt{In\-di\-ca\-tor\-Exas\-Vec\-tor}, \texttt{In\-di\-ca\-tor\-Exas\-Vec\-torSplit}, \texttt{API\-In\-di\-ca\-tor\-Exas\-Vec\-tor}, and \texttt{API\-In\-di\-ca\-tor\-Exas\-Vec\-torSplit} achieved a significantly better precision. 
In contrast, we could not determine significant differences within this group. 
Considering the recall, we can see that the \texttt{-Split-} functions have a significantly larger recall then the remaining functions. 
Within the \texttt{-Split-}-functions, we can observe that some \texttt{-Cosine}-functions have a significant larger recall than the \texttt{-L1Norm}-functions, for instance, \texttt{ExasVectorSplitCosine} compared to \texttt{ExasVectorSplitL1Norm}.

We also computed the effect sizes using Cliff's $\delta$~\cite{Hogarty1999,Kitchenham2017}. Our results indicate medium effect sizes for the conservative precision, while they are usually small for the recall.
However, the positive results regarding the precision may be a result of the large proportion of rules in the MUBench dataset originating from the project Joda-Time (i.e., 40 of 89). 
Nevertheless, the results are still valuable. 
First, they describe the applicability of our technique in a within-project setting (i.e., finding similar API misuses in the same project). 
Second, we identified individual differences between the distance functions to guide our other experiments.

We see that only a minority of misuses could be detected. Particularly, the mean numbers of absolute true positives detected by the best performing functions at threshold $<0.4$ are $\approx16$ for \texttt{In\-di\-ca\-tor\-Exas\-Vec\-tor}, $\approx18.4$ for \texttt{In\-di\-ca\-tor\-Exas\-Vec\-torSplit}, $\approx16$ for \texttt{API\-In\-di\-ca\-tor\-Exas\-Vec\-tor}, and $\approx19$ for \texttt{API\-In\-di\-ca\-tor\-Exas\-Vec\-torSplit}. Thus, we state that even though the functions achieve a high relative precision (i.e., low rate of `false alarms'), this technique is only a complement to existing API misuse detectors with a much higher recall.

The good performance is also indicated by the violin plot in \autoref{fig:violin_plot_dist}. This plot describes the distributions of the different use cases (particularly, for the \texttt{IndicatorExasVector} function) when comparing rules with actual API usages, such as the distance between the correct part (i.e., $rc$) of the rule and the misuse (i.e., $m$). We observe that the higher frequencies around $1$ and $0.5$ are almost equal in all cases. However, the frequency of distances towards the $0.0$ for the use case $dist(rc,m)$ is slightly larger than the others. This benefits the misuse detection conducted by \autoref{eq:misuse-detect} since then it is more likely that the distance of $dist(rc,m)$ is larger than the distance $dist(rm,m)$. Similarly, due to the equal distributions on the bottom of the violin plots, the distance of $dist(rc,c)$ is more unlikely to be larger than the distance $dist(rm,c)$. Thus, the misuse detection does not report correct usages as misuses. As discussed before, most distributions of the distance values are similarly shaped, which justifies the benefit of the applicability check. However, the number of applicable rules usually causes the different performance of the distance functions.

\myPar{Experiment 2: MUBench on AU500}
\begin{figure}
	\begin{subfigure}{.49\textwidth}
		\caption{Relative Precision\label{fig:rel_precision_au500}}
		\includegraphics[width=\textwidth, trim=58 58 70 100, clip]{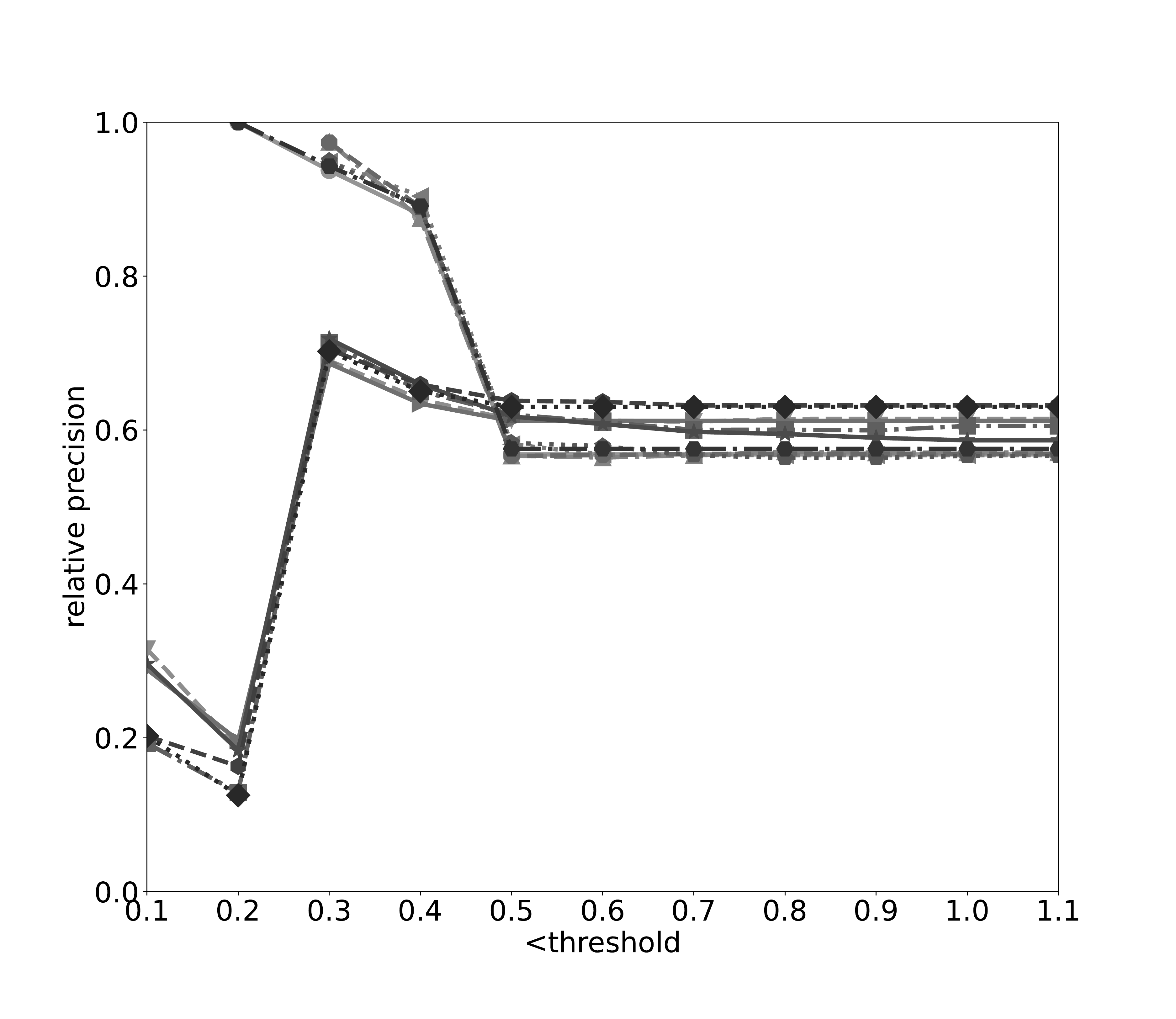}	
	\end{subfigure}	
	\begin{subfigure}{.49\textwidth}
		\caption{Conservative Precision\label{fig:cons_precision_au500}}
		\includegraphics[width=\textwidth, trim=58 58 70 100, clip]{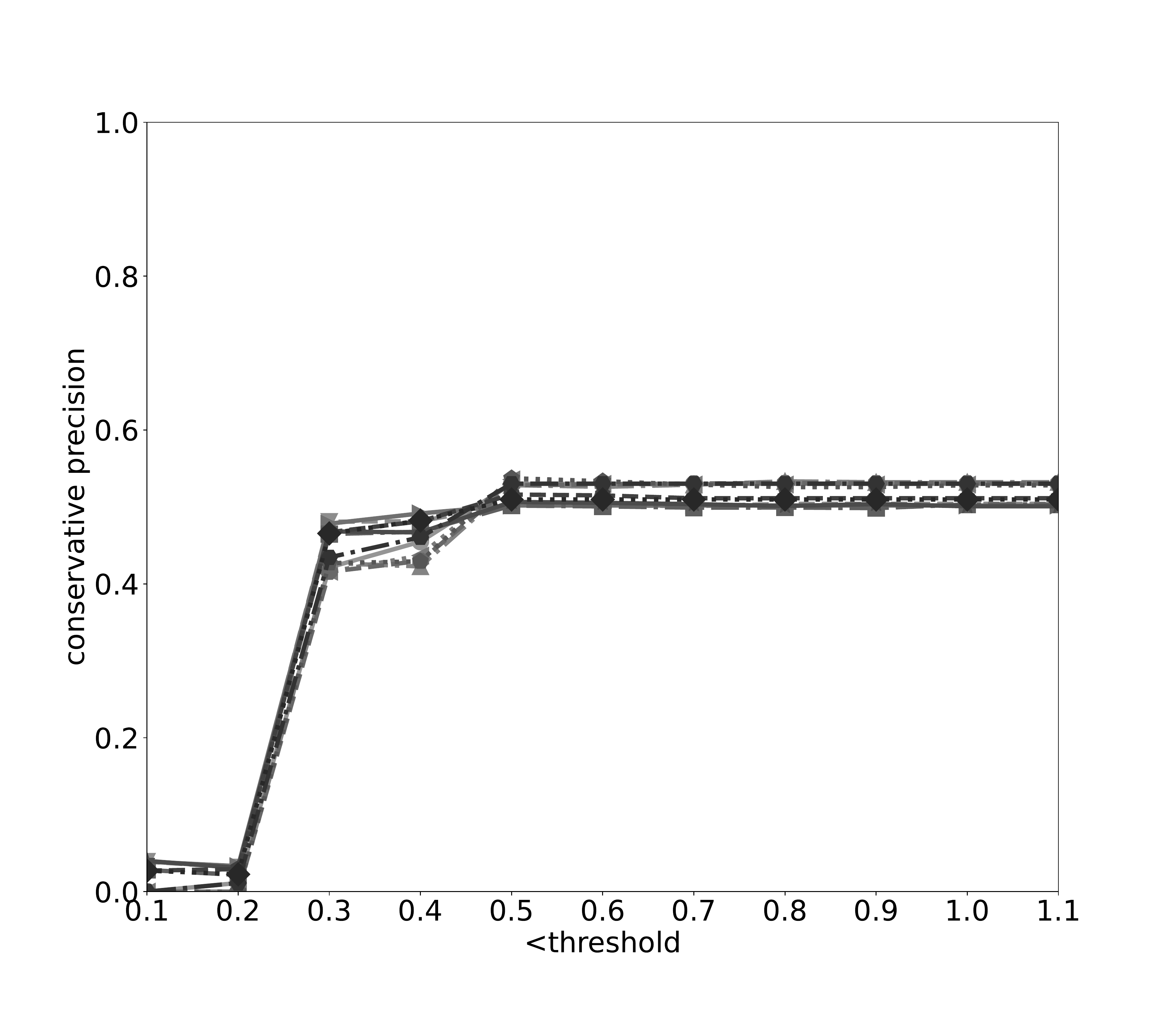}
	\end{subfigure}
	\begin{subfigure}{.49\textwidth}
		\caption{Recall\label{fig:recall_au500}}
		\includegraphics[width=\textwidth, trim=58 58 70 100, clip]{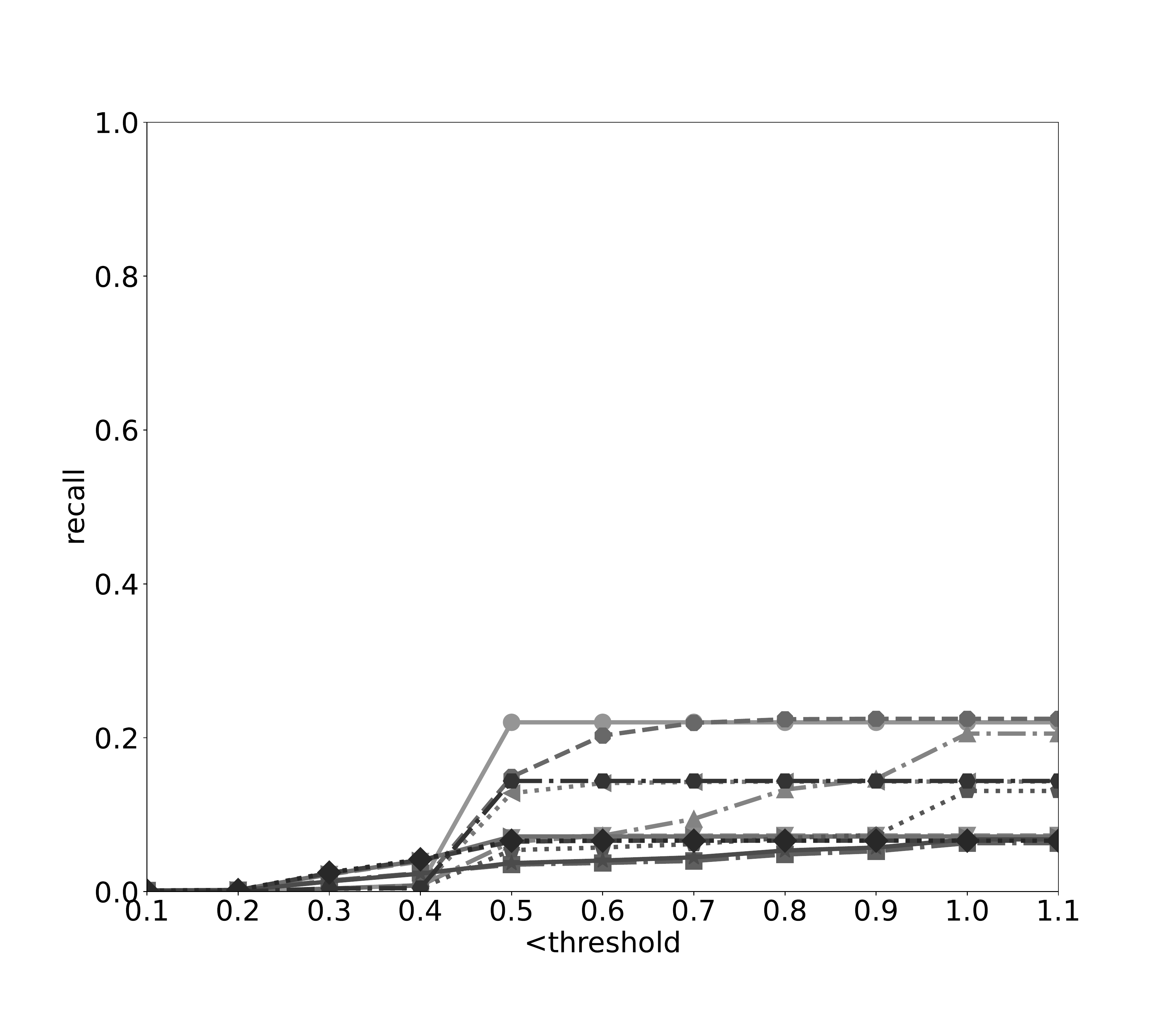}
	\end{subfigure}	
	\begin{subfigure}{\textwidth}
		\includegraphics[width=\textwidth, trim=70 140 70 50, clip]{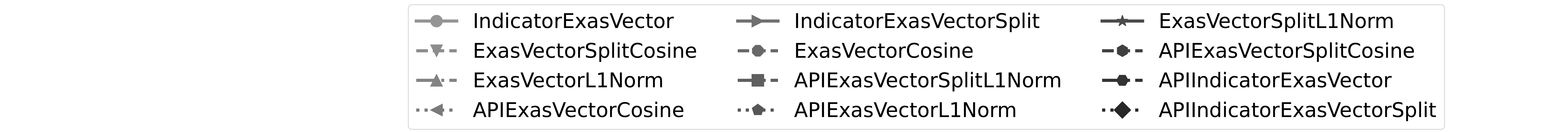}
	\end{subfigure}	
	\caption{Mean values of the performance values on AU500 dataset using different thresholds for applicability check \label{fig:misuse_au500}}
\end{figure}
\begin{table}
	\caption{Results MUBench on AU500 of the pairwise comparison of the conservative precision at threshold $<0.4$ using the Wilcoxon signed-rank test ($\alpha=0.05$) with Bonferroni correction. \cmark\ denotes the significant results while \xmark\ denotes the non-significant results. The $>$ or $<$ character denotes whether the mean conservative precision of the function on the left-hand side is $<$ or $>$ to the mean conservative precision of the function on the top. Numbers after significant results represent the effect size measured with Cliff's $\delta$.\label{tab:s_tests_prec_au500}}
	\begin{adjustbox}{max width=\linewidth}	
		\begin{tabular}{l|cccccccccccc}
			\toprule
			{} & \rotatebox{90}{APIExasVectorCosine} & \rotatebox{90}{APIExasVectorL1Norm} & \rotatebox{90}{APIExasVectorSplitCosine} & \rotatebox{90}{APIExasVectorSplitL1Norm} & \rotatebox{90}{APIIndicatorExasVector} & \rotatebox{90}{APIIndicatorExasVectorSplit} & \rotatebox{90}{ExasVectorCosine} & \rotatebox{90}{ExasVectorL1Norm} & \rotatebox{90}{ExasVectorSplitCosine} & \rotatebox{90}{ExasVectorSplitL1Norm} & \rotatebox{90}{IndicatorExasVector} & \rotatebox{90}{IndicatorExasVectorSplit} \\
			\midrule
			APIExasVectorCosine         &                   - &              \xmark &                   \xmark &                   \xmark &                 \xmark &                      \xmark &           \xmark &           \xmark &                \xmark &                \xmark &              \xmark &           < \cmark -0.13 \\
			APIExasVectorL1Norm         &              \xmark &                   - &           < \cmark -0.12 &                   \xmark &                 \xmark &              < \cmark -0.12 &           \xmark &           \xmark &        < \cmark -0.12 &                \xmark &              \xmark &           < \cmark -0.14 \\
			APIExasVectorSplitCosine    &              \xmark &       > \cmark 0.12 &                        - &                   \xmark &                 \xmark &                      \xmark &    > \cmark 0.12 &    > \cmark 0.13 &                \xmark &                \xmark &              \xmark &                   \xmark \\
			APIExasVectorSplitL1Norm    &              \xmark &              \xmark &                   \xmark &                        - &                 \xmark &                      \xmark &           \xmark &           \xmark &                \xmark &                \xmark &              \xmark &                   \xmark \\
			APIIndicatorExasVector      &              \xmark &              \xmark &                   \xmark &                   \xmark &                      - &                      \xmark &           \xmark &           \xmark &                \xmark &                \xmark &              \xmark &                   \xmark \\
			APIIndicatorExasVectorSplit &              \xmark &       > \cmark 0.12 &                   \xmark &                   \xmark &                 \xmark &                           - &    > \cmark 0.12 &    > \cmark 0.13 &                \xmark &                \xmark &              \xmark &                   \xmark \\
			ExasVectorCosine            &              \xmark &              \xmark &           < \cmark -0.12 &                   \xmark &                 \xmark &              < \cmark -0.12 &                - &           \xmark &        < \cmark -0.12 &                \xmark &              \xmark &           < \cmark -0.14 \\
			ExasVectorL1Norm            &              \xmark &              \xmark &           < \cmark -0.13 &                   \xmark &                 \xmark &              < \cmark -0.13 &           \xmark &                - &        < \cmark -0.13 &                \xmark &              \xmark &           < \cmark -0.15 \\
			ExasVectorSplitCosine       &              \xmark &       > \cmark 0.12 &                   \xmark &                   \xmark &                 \xmark &                      \xmark &    > \cmark 0.12 &    > \cmark 0.13 &                     - &                \xmark &              \xmark &                   \xmark \\
			ExasVectorSplitL1Norm       &              \xmark &              \xmark &                   \xmark &                   \xmark &                 \xmark &                      \xmark &           \xmark &           \xmark &                \xmark &                     - &              \xmark &                   \xmark \\
			IndicatorExasVector         &              \xmark &              \xmark &                   \xmark &                   \xmark &                 \xmark &                      \xmark &           \xmark &           \xmark &                \xmark &                \xmark &                   - &                   \xmark \\
			IndicatorExasVectorSplit    &       > \cmark 0.13 &       > \cmark 0.14 &                   \xmark &                   \xmark &                 \xmark &                      \xmark &    > \cmark 0.14 &    > \cmark 0.15 &                \xmark &                \xmark &              \xmark &                        - \\
			\bottomrule
		\end{tabular}
	\end{adjustbox}
\end{table}
\begin{table}
	\caption{Results MUBench on AU500 of the pairwise comparison of the recall at threshold $<0.4$ using the Wilcoxon signed-rank test ($\alpha=0.05$) with Bonferroni correction. \cmark\ denotes the significant results while \xmark\ denotes the non-significant results. The $>$ or $<$ character denotes whether the mean recall of the function on the left-hand side is $<$ or $>$ to the mean recall of the function on the top. Numbers after significant results represent the effect size measured with Cliff's $\delta$.\label{tab:s_tests_rec_au500}}
	\begin{adjustbox}{max width=\linewidth}	
		\begin{tabular}{l|cccccccccccc}
			\toprule
			{} & \rotatebox{90}{APIExasVectorCosine} & \rotatebox{90}{APIExasVectorL1Norm} & \rotatebox{90}{APIExasVectorSplitCosine} & \rotatebox{90}{APIExasVectorSplitL1Norm} & \rotatebox{90}{APIIndicatorExasVector} & \rotatebox{90}{APIIndicatorExasVectorSplit} & \rotatebox{90}{ExasVectorCosine} & \rotatebox{90}{ExasVectorL1Norm} & \rotatebox{90}{ExasVectorSplitCosine} & \rotatebox{90}{ExasVectorSplitL1Norm} & \rotatebox{90}{IndicatorExasVector} & \rotatebox{90}{IndicatorExasVectorSplit} \\
		\midrule
		APIExasVectorCosine         &                   - &                   - &           < \cmark -0.45 &           < \cmark -0.35 &                 \xmark &              < \cmark -0.47 &   < \cmark -0.18 &   < \cmark -0.17 &        < \cmark -0.44 &        < \cmark -0.35 &      < \cmark -0.22 &           < \cmark -0.47 \\
		APIExasVectorL1Norm         &                   - &                   - &           < \cmark -0.45 &           < \cmark -0.35 &                 \xmark &              < \cmark -0.47 &   < \cmark -0.18 &   < \cmark -0.17 &        < \cmark -0.44 &        < \cmark -0.35 &      < \cmark -0.22 &           < \cmark -0.47 \\
		APIExasVectorSplitCosine    &       > \cmark 0.45 &       > \cmark 0.45 &                        - &            > \cmark 0.08 &          > \cmark 0.43 &                      \xmark &           \xmark &           \xmark &                \xmark &         > \cmark 0.08 &              \xmark &                   \xmark \\
		APIExasVectorSplitL1Norm    &       > \cmark 0.35 &       > \cmark 0.35 &           < \cmark -0.08 &                        - &          > \cmark 0.33 &              < \cmark -0.11 &           \xmark &           \xmark &                \xmark &                \xmark &              \xmark &           < \cmark -0.11 \\
		APIIndicatorExasVector      &              \xmark &              \xmark &           < \cmark -0.43 &           < \cmark -0.33 &                      - &              < \cmark -0.46 &   < \cmark -0.16 &   < \cmark -0.15 &        < \cmark -0.42 &        < \cmark -0.34 &       < \cmark -0.2 &           < \cmark -0.45 \\
		APIIndicatorExasVectorSplit &       > \cmark 0.47 &       > \cmark 0.47 &                   \xmark &            > \cmark 0.11 &          > \cmark 0.46 &                           - &           \xmark &           \xmark &                \xmark &          > \cmark 0.1 &              \xmark &                   \xmark \\
		ExasVectorCosine            &       > \cmark 0.18 &       > \cmark 0.18 &                   \xmark &                   \xmark &          > \cmark 0.16 &                      \xmark &                - &           \xmark &                \xmark &                \xmark &              \xmark &                   \xmark \\
		ExasVectorL1Norm            &       > \cmark 0.17 &       > \cmark 0.17 &                   \xmark &                   \xmark &          > \cmark 0.15 &                      \xmark &           \xmark &                - &                \xmark &                \xmark &              \xmark &                   \xmark \\
		ExasVectorSplitCosine       &       > \cmark 0.44 &       > \cmark 0.44 &                   \xmark &                   \xmark &          > \cmark 0.42 &                      \xmark &           \xmark &           \xmark &                     - &         > \cmark 0.08 &              \xmark &                   \xmark \\
		ExasVectorSplitL1Norm       &       > \cmark 0.35 &       > \cmark 0.35 &           < \cmark -0.08 &                   \xmark &          > \cmark 0.34 &               < \cmark -0.1 &           \xmark &           \xmark &        < \cmark -0.08 &                     - &              \xmark &            < \cmark -0.1 \\
		IndicatorExasVector         &       > \cmark 0.22 &       > \cmark 0.22 &                   \xmark &                   \xmark &           > \cmark 0.2 &                      \xmark &           \xmark &           \xmark &                \xmark &                \xmark &                   - &                   \xmark \\
		IndicatorExasVectorSplit    &       > \cmark 0.47 &       > \cmark 0.47 &                   \xmark &            > \cmark 0.11 &          > \cmark 0.45 &                      \xmark &           \xmark &           \xmark &                \xmark &          > \cmark 0.1 &              \xmark &                        - \\
		\bottomrule
		\end{tabular}
	\end{adjustbox}
\end{table}
Similar to the previous experiment, we computed the mean values of the relative and conservative precision as well as the recall. The results are depicted in \autoref{fig:misuse_au500}. Again, we observe more dynamics in the mean values for a threshold $<0.5$ and lower, some fairly high increase in the mean relative precision of certain distance functions of more than $0.9$ between the thresholds $<0.2$ and $<0.4$ (cf. \autoref{fig:rel_precision_au500}) as well as a rapidly decreasing recall for lower thresholds (cf. \autoref{fig:recall_au500}). However, we observe that this drop, especially from $<0.5$ to $<0.4$, is steeper than in the MUBench on MUBench experiment. Therefore, we also consider the performance of the distance function at the threshold $<0.4$ since lower thresholds would only further decrease the recall without substantially increasing the relative precision.

At this threshold, the distance functions achieving the highest mean relative precision are \texttt{Exas\-Vec\-tor\-L1Norm} ($\approx87.4\%$),  \texttt{Exas\-Vec\-tor\-Co\-sine} ($\approx88.9\%$), \texttt{API\-Exas\-Vec\-tor\-L1Norm} ($\approx88.8\%$), \texttt{API\-Exas\-Vec\-tor\-Co\-sine} ($\approx90.4\%$), \texttt{In\-di\-ca\-tor\-Exas\-Vec\-tor} ($\approx88\%$), and \texttt{API\-In\-di\-ca\-tor\-Exas\-Vec\-tor} ($\approx89.1\%$). Regarding the conservative precisions (cf. \autoref{fig:cons_precision_au500}) and the recall (cf. \autoref{fig:recall_au500}) at threshold $<0.4$ the visual differences are subtle. Nevertheless, we pairwise checked whether the differences are significant (cf. \autoref{tab:s_tests_prec_au500} and \autoref{tab:s_tests_rec_au500}).

We observe that the distance functions \texttt{API\-Exas\-Vec\-tor\-Split\-Co\-sine},  \texttt{API\-In\-di\-ca\-tor\-Exas\-Vec\-tor\-Split}, \texttt{Exas\-Vec\-tor\-Split\-Co\-sine}, and \texttt{In\-di\-ca\-tor\-Exas\-Vec\-tor\-Split} have a significant better conservative precision than mainly the non-\texttt{-Split-}-functions. However, the effect size is usually small for these cases. Regarding the recall, we observe that mainly the \texttt{-Split-}-functions have some significantly higher recall than other functions with partially large effect sizes, such as \texttt{In\-di\-ca\-tor\-Exas\-Vec\-tor\-Split} compared to \texttt{API\-Exas\-Vec\-tor\-Co\-sine}.

In comparison to the results of the first experiment, we also observe the same positive effect on the increase of relative precision when decreasing the threshold. In this setting the \texttt{-Split-}-functions perform better even though the effect is less strong. However, the recall is usually very low. For instance, the mean number of absolute true positives at threshold $<0.4$ ranges from $0.4$ up to $4.7$, while we tested for $114$ misuses. Therefore, this cross-project setting is less successful in detecting the majority of misuses. 
Nevertheless, if misuses are detected the chance of `false alarms' is usually very low.

\begin{figure}
	\begin{subfigure}{\textwidth}
		\caption{Relative Precision\label{fig:rel_precision_androidcompass}}
		\includegraphics[width=\textwidth]{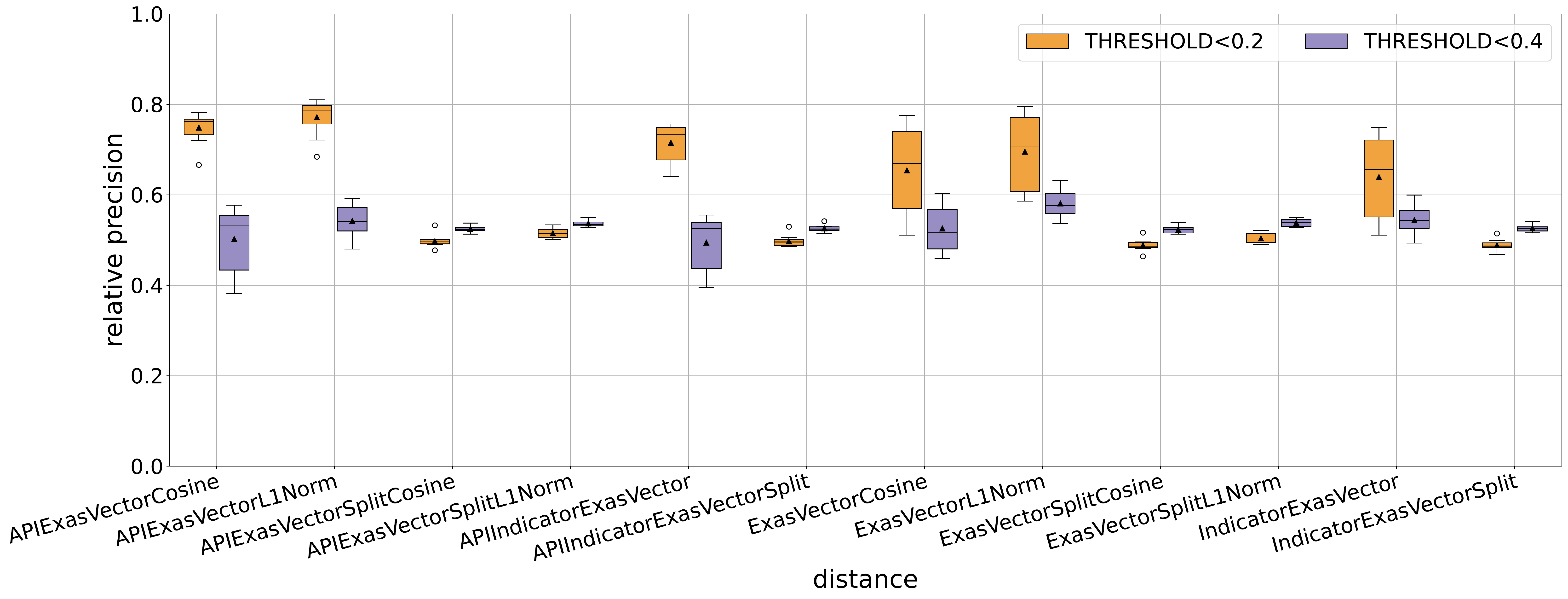}	
	\end{subfigure}	
	\begin{subfigure}{\textwidth}
		\caption{Conservative Precision\label{fig:cons_precision_androidcompass}}
		\includegraphics[width=\textwidth]{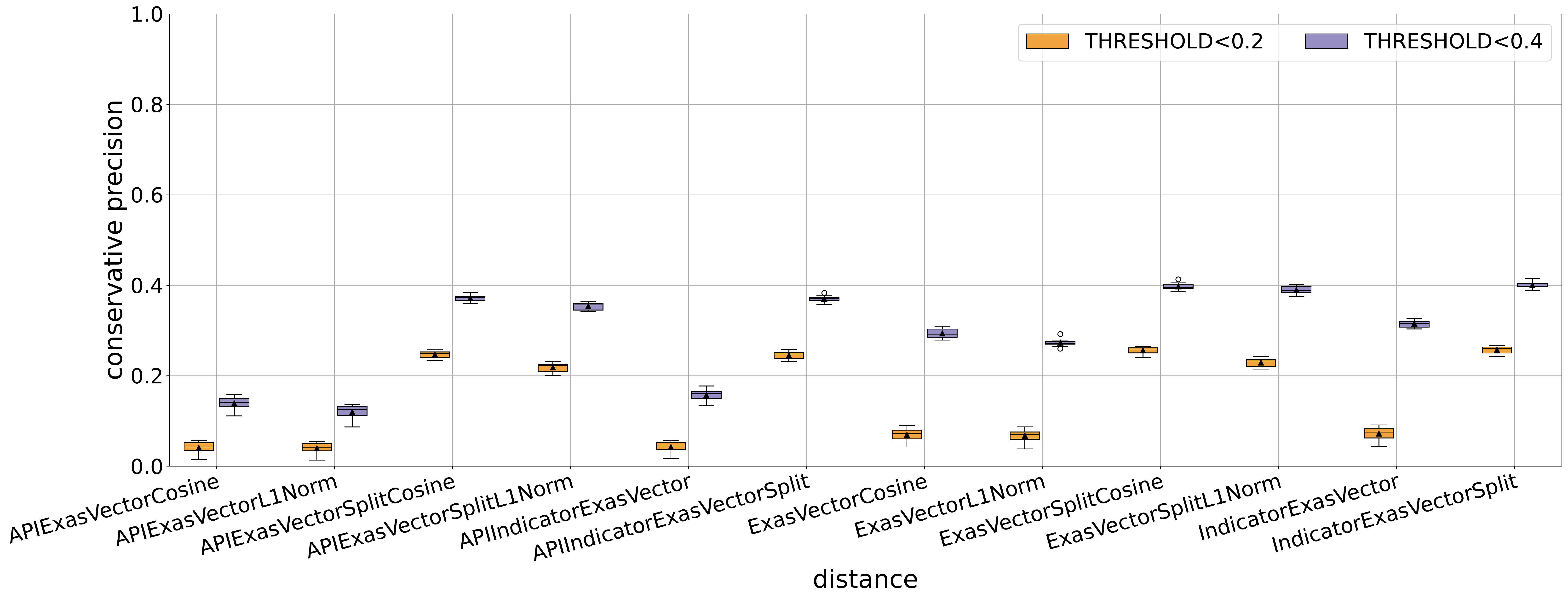}
	\end{subfigure}
	\begin{subfigure}{\textwidth}
		\caption{Recall\label{fig:recall_androidcompass}}
		\includegraphics[width=\textwidth]{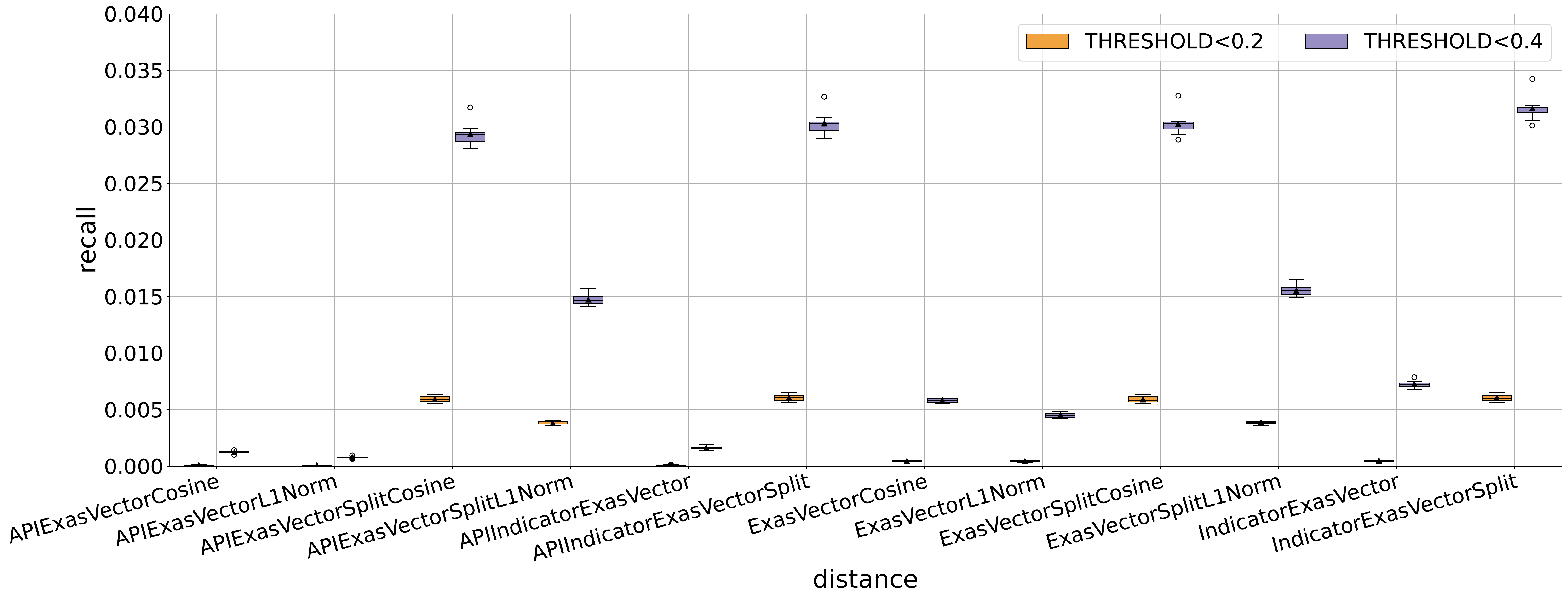}
	\end{subfigure}	
	\caption{Boxplot of the mean values of the performance values on the AndroidCompass dataset using different thresholds $<0.2$ and $<0.4$ for applicability check. Due to the low recall values, the ordinate of the recall plot (subfigure~\ref{fig:recall_androidcompass} is rescaled). \label{fig:misuse_androidcompass}}
\end{figure}

\myPar{Experiment 3: AndroidCompass on AndroidCompass}
Due to the huge amount of distance computations in this setting, we ran the experiments on a compute cluster (processor 64x 2.9 GHz with hyper-threading, 1TB RAM). The computation of all twelve distance values among all ten-fold cross-validation subsets took approximately three weeks on this cluster. We computed the performance values (i.e., relative and conservative precision as well as recall) as described in Section~\ref{ssec:datasets}. To reduce the number of computations we only considered the threshold-values of $0.2$ and $0.4$ based on our experience from the previous two experiments. The distributions of the performance values are depicted as boxplot in \autoref{fig:misuse_androidcompass}.

In detail, the best mean values of the relative precision were achieved with $THRESHOLD$-value $0.2$ by \texttt{API\-Exas\-Vec\-tor\-L1Norm} ($\approx77.1\%$), \texttt{API\-Exas\-Vec\-tor\-Co\-sine} ($\approx74.8\%$), and \texttt{API\-In\-di\-ca\-tor\-Exas\-Vec\-tor} ($\approx71.4\%$). This is followed by \texttt{Exas\-Vec\-tor\-L1Norm} ($\approx69.5\%$), \texttt{Exas\-Vec\-tor\-Co\-sine} ($\approx65.3\%$), and \texttt{In\-di\-ca\-tor\-Exas\-Vec\-tor} ($\approx63.9\%$). Note that these values have a larger variance as indicated by \autoref{fig:rel_precision_androidcompass}.  The \texttt{-Split-} distances usually perform worse regarding the relative precision, however, they show a larger conservative precision with $THRESHOLD$ $0.4$ (cf. \autoref{fig:cons_precision_androidcompass}) than the non-\texttt{-Split-} variants. Except for the \texttt{-Split-} distances, we can also observe an improved relative precision when using the smaller $THRESHOLD$ $0.2$. The mean recall is very close to zero (cf. Figure~\ref{fig:recall_androidcompass}) ranging from $0.0074\%$ for \texttt{API\-Exas\-Vec\-tor\-L1Norm} at $THRESHOLD$ $0.2$ to $3.1\%$ for \texttt{In\-di\-ca\-tor\-Exas\-Vec\-tor\-Split} at $THRESHOLD$ $0.4$.

Again, we observed the positive effect on the relative precision when decreasing the $THRESHOLD$ value (except for \texttt{-Split-} distances). Regarding the relative precision distances using Exas vectors with API-specific features perform best, while regarding the conservative precision distances using the \texttt{-Split-}-feature perform better. This is similar to the second experiment and may indicate that these distances perform best in a cross-project setting. Again, the recall is very low, manifesting the insight that the distance-based misuse detection may work best in a hybrid setting together with other misuse detection mechanisms (e.g., applying frequent API usage patterns).

\section{Discussion\label{sec:discussion}}

\subsection{Implications of the Results and Comparison to other API Misuse Detectors}

In this section, we compare the results of \tool with its respective distance-based misuse detection to the performance values, particularly, precision and recall, of other API misuse detectors. We collected these state-of-the-art tools based on the works by Amann et al.~\cite{Amann2018a, Amann2019} and Kang and Lo~\cite{Kang2021} as well as our own literature research applying a forward snowballing-like approach based on these publications. An overview of the comparison is depicted in \autoref{tab:comparison_rw}. We emphasize that the results are based on the reports of the respective papers and apart from those not directly applied to the MUBench or AU500 dataset are only indirectly comparable. Nevertheless, this shows the current state of the performance of API misuse detectors and whether our results represent an improvement regarding the current state. Moreover, our approach measures the mean precision and recall over the values achieved by the \emph{single, applicable rules}. Thus, it is very unlikely to find values for those rules providing a high recall, since we do not expect to have a `super-rule' detecting a majority of misuses.

\begin{table}
	\caption{Performance values of state-of-the-art misuse detectors (N/A = not available)\label{tab:comparison_rw}}
	\begin{minipage}{\textwidth}
	\begin{tabular}{llrr}
		\hline
		\textbf{Detector} & \textbf{Validation Datasets} & \textbf{Precision} & \textbf{Recall}\\\hline
		ALP~\cite{Kang2021} & \textit{\textbf{AU500}}, \textit{\textbf{MUBench}} & $43.9-44.7\%$ & $54.8-56.3\%$\\\hline
		MUDetect~\cite{Amann2019a,Kang2021} & \textit{\textbf{AU500}}, \textit{\textbf{MUBench}} & $27.6-34.1\%$ & $29.6-43.3\%$\\\hline
		Ren et al.~\cite{Ren2020} & \textit{\textbf{MUBench}} & $60.2\%$ & $28.45\%$\\\hline
		FuzzyCatch~\cite{Nguyen2020} & self-collected & $65-92\%$ & $73.4-82.1\%$\footnote{recall on manually selected exception bugs} \\\hline
		Salento~\cite{Murali2017} & self-collected & $75\%$ & $100\%$\footnote{no specific numbers for true/false positive/negatives reported} \\\hline
		Pradel et al.~\cite{Pradel2012} & DeCapo\cite{Blackburn2006} & $50.6\%$ & $70\%$\\\hline
		Pradel and Gross~\cite{Pradel2012a} & DeCapo & $100\%$ & N/A\\\hline
		Tikanga~\cite{Wasylkowski2011,Amann2019} & self-collected, \textit{\textbf{MUBench}} & $11.4-39.7\%$ & $13.2\%$\footnote{only available for MUBench \label{fn:mubench_avail}}\\\hline
		SpecCheck~\cite{Nguyen2011} & DeCapo & $54.2\%$ & N/A \\\hline
	    DMMC~\cite{Monperrus2010, Amann2019} & self-collected, \textit{\textbf{MUBench}} & $9.9-57.9\%$ & $20.8\%$\textsuperscript{\ref{fn:mubench_avail}}\\\hline
		OCD~\cite{Gabel2010} & self-collected & $60\%$ & N/A \\\hline
		GROUMiner~\cite{Nguyen2009a, Amann2019} & self-collected, \textit{\textbf{MUBench}} & $0-13.9\%$ & $0\%$\textsuperscript{\ref{fn:mubench_avail}}\\\hline
		CAR-Miner~\cite{Thummalapenta2009a} & subset from~\cite{Weimer2005}& $64.3\%$ & $80\%$\footnote{relative recall in comparison to WN-miner specifications~\cite{Weimer2005}} \\\hline
		Acharya and Xie~\cite{Acharya2009} & self-collected & $90.4\%$ & N/A \\\hline
		Alattin~\cite{Thummalapenta2009} & self-collected & $37.8\%$ & $94.9\%$\footnote{relative recall in comparison to a baseline approach}\\\hline
		Jadet~\cite{Wasylkowski2007, Amann2019} & self-collected, \textit{\textbf{MUBench}} & $10.3-48.1\%$& $5.7\%$\textsuperscript{\ref{fn:mubench_avail}}\\\hline
		PR-Miner~\cite{Li2005} & self-collected & $23.5\%$ &  N/A \\\hline
		\textbf{Our Approach} & \textit{\textbf{MUBench}}, \textit{\textbf{AU500}}, \textit{\textbf{AndroidCompass}} & \textbf{$77.1-96.1\%$} & \textbf{$0.007-17.7\%$} \\\hline
	\end{tabular}
	\end{minipage}
\end{table}

\myPar{Misuse Detectors on the same dataset}
One very recent and related result is based on the work by Kang and Lo with their active-learning-based approach ALP~\cite{Kang2021}. In their work, they compare ALP with the cross-project version of MUDetect~\cite{Amann2018a,Amann2019a}, namely, MUDetectXP on the datasets MUBench and AU500. Compared with our results, their recall is larger, however, their precision is much lower. 

Amann et al. tested Tikanga~\cite{Wasylkowski2011}, DMMC~\cite{Monperrus2010}, GROUMiner~\cite{Nguyen2009a}, and Jadet~\cite{Wasylkowski2007} on the MUBench dataset and also stated a much lower precision than ours and comparable or partially better recall for MUBench (i.e., our value was $17.7\%$ for MUBench). For the tested approaches, even on the self-collected datasets of those detectors the precision was lower than seen in our results.

Ren et al.~\cite{Ren2020} built a detector based on a knowledge graph constructed from caveats of the API documentation and tested them on a subset of MUBench. They measured the precision by checking the code version containing the misuse with their tooling and assessing whether the explanation matches the misuse description. They do not report the performance on real negative results, particularly, the false positive rate for code not containing an API misuse. Similar to the work by Kang and Lo, they achieved a higher recall than our approach. In opposite, we do not require information from the documentation.

\myPar{Misuse Detectors on other datasets}
A very promising and recent approach is FuzzyCatch which achieves fairly good precision (i.e., comparable to our results) and recall values on their self-collected dataset. However, their approach is designed and tested to detect missing exception handling procedures, while our approach is not particularly limited to a certain kind of misuse as long as there exists a fixing commit. Similarly, the approach by Acharya and Xie~\cite{Acharya2009} considers error handling specifications in C and could also achieve comparable results. 

Salento~\cite{Murali2017} is a tool for learning a statistical model using a Bayesian framework to discriminate misuses from correct usages. While achieving promising results, it was currently tested only on Android apps and their tool requires a learning phase and data. In our work, we only need a fixing commit to trigger a misuse detection.

Pradel and Gross~\cite{Pradel2012a} developed a misuse detector inferring patterns from execution traces of automatically generated test runs. In their experiment, they achieved a zero false-positive rate. However, this is a dynamic approach since the pattern inference requires the execution of the code. This is not necessary for \tool and the subsequent misuse detection. Similarly, the approach by Pradel et al.~\cite{Pradel2012} infers multi-object specifications by a dynamic specification miner.

Finally, the tools SpecCheck~\cite{Nguyen2011}, OCD~\cite{Gabel2010}, CAR-miner~\cite{Thummalapenta2009a}, Alattin~\cite{Thummalapenta2009}, and PR-Miner~\cite{Li2005} performed worse regarding the precision while the recall was either not reported or only compared to a baseline approach and reported as the relative recall, i.e., the proportion of detected misuses found by any of the tested detectors. 

\myPar{Summary of the Comparison}
We found that \tool and its misuse detection is a promising approach to increase the precision of API misuse detection. However, regarding the recall, our approach performs worse than other detectors. This is caused by our measurement of the performance which focuses on single rules instead of a set of multiple rules. For the latter, we require a technique to select and rank rules in case multiple rules were applicable. Based on our results this ranking could be done along the $dist(aug_{rm},aug_{x})$-value (cf. \autoref{eq:applic}). Having a strong cut-off of applicable rules (i.e., the $THRESHOLD$-value), we expect to keep the observed large precision while increasing the recall. This needs to be tested in future analyses.

Moreover, \tool is a static approach (i.e., no code execution needed) and does not require a large-scale pattern mining, since it relies only on previously fixed misuses. Indeed, we require a proper set of fixes to infer change rules. However, similarly, pattern mining-based approaches require a set of high-quality code samples to infer patterns from. Due to its simplicity (i.e., comparing API usages via vectors distances), we expect it to be scalable. In \autoref{sec:conclusion}, we discuss some potential improvements for speeding up the vector generation and distance computation.

\subsection{Threats to Validity}

In this section, we discuss potential threats to the validity of our experimental results. As discussed by Wohlin et al.~\cite{Wohlin2012} (pp. 104ff), we consider threats to the \textit{internal}, \textit{external}, and \textit{construct} validity.

\myPar{Internal Validity}
This paragraph discusses threats that may have influence on the independent values and thus may bias the drawn conclusions. 

The results of our preliminary study on the quality of produced correction rules may be biased due to the subjective assessment and preference of the first three authors. Another independent group may validate these rules differently or would suggest different solutions. Thus, the results are part our replication package\textsuperscript{\ref{fn:replication}}.

We only compared our results to other detectors based on their reports. However, even though some used the same dataset, we did not have control over those experiments. For instance, some detectors may only be validated on a subset, which we did not consider due to our previous filtering. Thus, the comparison of the precision values may not directly match due to different data, which would explain the differences in the precision values.

In our experiments, we implemented several timeout mechanisms to cope with issues during git checkouts or long-lasting analyses. This may have an effect on the results, particularly, for smaller datasets, such as in the MUBench on MUBench experiment. For instance, when code samples were excluded, which would have produced bad performing change rules, these would bias the results.

In the cross-project setting, particularly, in the AndroidCompass dataset, repositories with an equal history (i.e., forked projects) may exist in different buckets of the cross-fold validation. This may bias the misuse detection, since then the misuse could have been detected by effectively the same change. We mitigated this effect by using cross-fold validation. Nevertheless, for re-validation purposes, we provide data and scripts in our replication package.

We did not control whether the fixing commit was present at the time the misuse existed. That means at the time the misuse was present, it was not clear whether we could construct the change rule that detects that misuse.

Moreover, the high precision may also be caused by detecting only limited types of misuses while not detecting other kinds, which could explain the low recall. We tried to mitigate this by applying different and independent datasets.

\myPar{External Validity}
This paragraph discusses threats hampering the generalizability of our results. 

The MUBench on MUBench experiment is considered as a project-internal setting since many misuses came from a single project (i.e., Joda-time). This may have only limited expressiveness for other project-internal settings. It may perform worse for projects without matching fixes in the version history. Nevertheless, the cross-project setting still achieves good precision values.

Our approach is designed for the object-oriented programming language Java and particularly the AUG generation leverages some details of its syntax. Thus, we cannot ensure whether our results apply to other programming languages and program paradigms. To apply \tool and its subsequent distance-based misuse detection, one must implement a similar transformation to AUGs and test its ability to detect misuse by using proper datasets.

\myPar{Construct Validity}
This paragraph considers potential issues with experimental design and the measurements, namely, whether the values are representative of the desired effect.

Single rule performance (i.e., precision, recall) may have only limited expressiveness, for instance, when multiple rules are applicable.
For that purpose, we require ranking strategies for rules similar to a pattern-based ranking. Depending on the strategy, the precision may differ from our results. Having a good cut-off criterion (i.e., $THRESHOLD$-value) and a high average precision among the applicable rules, we hope that this threat is mitigated. This will be validated in future work.

We only assume that the changing commit within the AndroidCompass dataset represents a misuse fix. In case this is not true, namely, the version before the commit is no misuse, our conclusion does not refer to misuse detection but rather detection of upcoming changes.

We do not report the severity of detected misuses. This may be important for the developers since they likely want to fix severe security bugs before handling inconvenient GUI errors.
\section{Related Work\label{sec:relatedwork}}

Next to works on API misuse detectors, our work relates to other domains, particularly, historical and collaborative bug detection and program repair, as well as research on code matching metrics.

\subsection{Historical or Context-related Bug Detection and Program Repair}

There is a huge domain working on the topic of automated program repair (APR)~\cite{Monperrus2018}. These tools naturally also apply bug detection and localization techniques, and thus, partially relate to our topic of misuse detection. However, mainly they do not consider API misuses in particular and their goal is to fix a bug eventually.

Among these approaches, several ideas were applied to leverage the information from past fixes (i.e., historical) to overcome bugs in similar code fragments (i.e., context-related). For instance, tools may use code-similarities to rank or retrieve patch candidates~\cite{Le2016, Xin2017}. Very recently, tools applied mining and deep-learning-based technologies to learn patterns from previous or context-related fixes and achieved good results in comparison to standard APR tools~\cite{Tufano2019, Bader2019, Kim2019, Li2020, Koyuncu2020, Liu2021}. In opposite to \tool and our misuse detection, these tools require separate data and a training phase to fix bugs, while we aim to transfer specific fixes to other locations.

Many program repair tools apply automated tests to identify code sites that must be altered to fix a bug. Usually, these approaches mark lines (i.e., spectra) that are more frequently executed when applying failing than passing tests, which is denoted as spectrum-based fault localization (SBFL). Wen et al. introduced a historical SBFL technique that automatically ranks located bugs by the ratio of bug-inducing and non-bug-inducing commits which add these lines~\cite{Wen2021}. However, SBFL only locates faults, while our approach also provides possible solutions in the form of change rules.

Nevertheless, these approaches also demonstrate the beneficial impact on bug detection and repair when applying information from other, similar fixes.

\subsection{Cooperative Bug Detection and Program Repair}

Cooperative bug detection and program repair aim to support developers in finding and fixing bugs by interacting with other developers or their code and leveraging their previous knowledge. 

Liu et al. introduced SOFix, an APR approach, which leverages code samples from StackOverflow discussions to infer repair templates~\cite{Liu2018}. However, their tool lacks the interactive component, since their templates are only inspired by the extracted code samples.

Tan and Li introduced Bugine for Android apps that applies natural language processing and a ranking technique to find related issue reports on GitHub. They found it supportive to identify possible test scenarios and thus bugs, for student developers of Android apps~\cite{Tan2020}. 

\subsection{Code Matching Techniques}

In our work, we compare code changes with potential other misuses. This is a related task in code search and code clone detection. 

For instance, in our previous work on commit-based preprocessing to improve the pattern quality, we applied a keyword-based approach finding similar API usages~\cite{Nielebock2021a} which was inspired by similar code search tools~\cite{Holmes2005, Sahavechaphan2006}. In our prior work~\cite{Nielebock2021a}, we also discussed related code search techniques which apply domain-specific languages~\cite{Paul1994}, used code elements such as test cases, method declarations, or input/output examples to retrieve code samples~\cite{Lemos2007, Reiss2009, Stolee2014}, as well as similarities in the documentation~\cite{Kim2010}. Gu et al. applied a deep learning technique to infer embeddings of code to relate similar code~\cite{Gu2018}. 

In the domain of code clone detection techniques range from detecting syntactical equal to semantically (but not necessarily syntactically) equal code~\cite{Koschke2007, Roy2009}. The well-known \texttt{Deckard} tool used an AST-vectorization and thus inspired the Exas Vector technique~\cite{Jiang2007}. Modern code clone detectors apply deep learning techniques to relate semantic code clones~\cite{White2016, Saini2018} or aim to scale clone detection for large code bases~\cite{Sajnani2016}. 

Both code search and clone detection are only loosely related to our approach since those tools usually do not report the similarity metrics but rather return similar code samples.
\section{Conclusion\label{sec:conclusion}}

In this paper, we considered API misuses, namely, the false application of API elements from libraries by developers. In the past several automated misuses detectors were reported. However, many of them have a high false-positive rate. 

Thus, we propose \tool with a subsequent distance-based API misuse detector. Its main idea is that developers, who already fixed a misuse, provide the change information of the fix to detect similar kinds of misuses in other code sites or projects. For that purpose, \tool infers so-called change rules from the manually marked fixing commit. These rules describe the edit operations conducted between the misuse and fix version, represented as API usage graphs (AUGs), a data structure developed by Amann et al.~\cite{Amann2019a}. With this change rule, our approach conducts a distance-based applicability check and a misuse detection. Particularly, the distances were computed on a vector representation named Exas Vector, suggested by Nguyen et al.~\cite{Nguyen2009}.

We evaluated our approach with three experiments by employing three different data sets, namely, MUBench, AU500, and AndroidCompass. In these experiments, we evaluated different revised distance functions. Our experiments reveal that our revisions have significant positive effects on the relative precision (i.e., the proportion of applicable rules correctly identifying misuses) by achieving values between 77.1\,\%-96.1\,\%. This is a major improvement compared to other static misuse detectors. Particularly, we identified that having a strong applicability condition (i.e., low $THRESHOLD$-value) as well as distance functions considering mainly API-specific features (i.e., \texttt{API-}) and the presence of features instead of their frequencies (i.e., \texttt{Indicator-}) have positive effects. Currently, our evaluation only considers the performance of single rules, and thus, only achieves a low recall (i.e., between 0.007\,\%-17.7\,\%). Therefore, we currently consider \tool and its distance-based misuse detector as a precise complement to existing API misuse detectors.

In future work, we want to further improve and investigate the applicability of \tool and the distance-based misuse detection.

\myPar{Improving Recall} To increase the recall we need to employ a larger and more diverse number of applicable rules. This requires a selection and ranking mechanism to decide which rule should be applied for detection. In our experiments, we observed that the distance between the misuse part of the rule and the API usage under test could serve as such a ranking criterion. In further work, we will evaluate whether such a ranking can improve the recall without harming the high precision.

\myPar{Scaling} \tool and the distance-based misuse detection were designed to scale to large repository storing systems such as GitHub, for instance, as a service for their users. We observed, especially, in the \textit{AndroidCompass on AndroidCompass} experiment a long runtime. We assume that this is due to the non-optimized implementation, which was mainly designed to check effectiveness (i.e., reducing false-positive rate) rather than improving efficiency. In the future, we want to optimize the vectorization of AUGs and the distance computation by transforming it into matrix operations that can be faster executed on GPUs. For instance, to compute \emph{n}-paths features of Exas Vector one can perform multiple adjacency matrix multiplications or one can summarize the distance computation of a set of Exas Vectors by applying linear matrix operations.

\myPar{API Misuse Repair} While we currently only detect misuses, our vision is to also repair them automatically, namely, developing an automated program repair (APR) tool. As recent research has denoted~\cite{Kechagia2021}, existing APR tools can not directly be applied to API misuses, and thus specific API misuse repair tools have to be developed. We envision automatically applying the edits described by the change rule to the code at hand to produce patches. One open issue is the validation of these patches. General APR tools usually employ test suites. Since we do not require such test suites, we currently rely on a human validation of the generated patches.

\bibliographystyle{ACM-Reference-Format}
\bibliography{publishers,fullBib,references,MYfull,misuse-detection-fixing-rules} 
\end{document}